\documentclass[a4paper,11pt]{article}
\pdfoutput=1 

\usepackage{jcappub} 
\usepackage{aas_macros}
\usepackage[utf8]{inputenc}
\usepackage{graphicx,color,amssymb,amsmath,mathtools,cases,bm,float,tocbibind, subfig}
\usepackage{enumitem}
\usepackage[format=hang,font=small,textfont=it]{caption}

\title{Towards ending the partial sky E-B ambiguity in CMB observations}

\author[a,b,1]{Shamik Ghosh,\note{Corresponding author}}
\author[c,d,e]{Jacques Delabrouille,}
\author[a,b]{Wen Zhao,}
\author[f]{and Larissa Santos}

\affiliation[a]{CAS Key Laboratory for Research in Galaxies and Cosmology, Department of Astronomy, University of Science and Technology of China, Hefei 230026, China}
\affiliation[b]{School of Astronomy and Space Sciences, University of Science and Technology of China, Hefei, 230026, China}
\affiliation[c]{Laboratoire Astroparticule et Cosmologie (APC), CNRS/IN2P3, Universit\'e Paris Diderot, 75205 Paris Cedex 13, France}
\affiliation[d]{IRFU, CEA, Universit\'e Paris Saclay, 91191 Gif-sur-Yvette, France}
\affiliation[e]{Department of Astronomy, School of Physical Sciences, University of
Science and Technology of China, Hefei, Anhui 230026, China}
\affiliation[f]{Center for Gravitation and Cosmology, College of Physical Science and Technology, Yangzhou University, Yangzhou, 225009, China}

\emailAdd{shamik@ustc.edu.cn}
\emailAdd{delabrouille@apc.in2p3.fr}
\emailAdd{wzhao7@ustc.edu.cn}
\emailAdd{larissa@yzu.edu.cn}

\abstract{A crucial problem for partial sky analysis of CMB polarization is the $E$-$B$ leakage problem. Such leakage arises from the presence of `ambiguous' modes that satisfy properties of both $E$ and $B$ modes. Solving this problem is critical for primordial polarization $B$ mode detection in partial sky CMB polarization experiments. In this work we introduce a new method for reducing the leakage. We demonstrate that if we complement the $E$-mode information outside the observation patch with ancillary data from full-sky CMB observations, we can reduce and even effectively remove the $E$-to-$B$ leakage. For this objective, we produce $E$-mode Stokes $QU$ maps from Wiener filtered full-sky intensity and polarization CMB observations. We use these maps to fill the sky region that is not observed by the ground-based experiment of interest, and thus complement the partial sky Stokes $QU$ maps. Since the $E$-mode information is now available on the full sky we see a significant reduction in the $E$-to-$B$ leakage. We evaluate on simulated data sets the performance of our method for a `shallow' $f_\text{sky}=8\%$, and a `deep' $f_\text{sky}=2\%$ northern hemisphere sky patch, with AliCPT-like properties, and a LSPE-like $f_\text{sky}=30\%$ sky patch, by combining those observations with Planck-like full sky polarization maps. We find that our method outperforms the standard and the pure-$B$ method pseudo-$C_\ell$ estimators for all of our simulations. Our new method gives unbiased estimates of the $B$-mode power spectrum through-out the entire multipole range with near-optimal pseudo-$C_\ell$ errors for $\ell>20$. We also study the application of our method to the CMB-S4 experiment combined with LiteBIRD-like full sky data, and show that using signal-dominated full sky $E$-mode data we can eliminate the $E$-to-$B$ leakage problem.}

\begin{document}
\maketitle
\flushbottom

\section{Introduction}
\label{sec:intro}

After the final publication of the Planck space mission Cosmic Microwave Background (CMB) observations \citep{2018arXiv180706205P}, much of the attention of the CMB community has turned towards the precise measurement of CMB polarization anisotropies, and in particular the detection of primordial polarization patterns originating from inflationary gravitational waves.

CMB polarization on the celestial sphere can be decomposed in two distinct contributions with different properties \citep{1997PhRvD..55.1830Z,1997PhRvD..55.7368K}. $E$ modes, of even parity, are generated by all types of primordial perturbations of the spacetime metric. Plasma motions at last scattering, primarily due to acoustic oscillations generated by the time-evolution of primordial scalar (density) perturbations, are the main source of CMB polarization $E$ modes. Polarization $B$ modes, of odd parity, and of much lower amplitude than polarization $E$ modes, are not directly generated by scalar perturbations. They mostly arise after last scattering, from distortions of the polarization $E$-mode  pattern by gravitational lensing along the photon path. They can also be generated in the early universe by tensor perturbations of the metric (gravitational waves). The detection of those primordial gravitational waves is essential to understand the physics at work in the early universe, at energy scales comparable to the Planck scale, and in particular, to constrain models of cosmic inflation \citep{2016ARA&A..54..227K}.

Future space mission concepts for measuring CMB polarization have been proposed or are currently under study~\citep{2016SPIE.9904E..0WK,2018JCAP...04..014D,2019JLTP..194..443H,2019arXiv190210541H,2019arXiv190901591D}, but none of them is expected to be launched before the end of the 2020s. In the mean time, progress in the CMB field relies on a programme of sub-orbital experiments that will observe limited regions of the celestial sphere. By reason of contamination of CMB observations by foreground astrophysical emission, even those experiments capable of observing substantial fractions of sky must limit cosmological analyses to clean regions of typically a few percent to a few tens of percents of the total sky. The ultimate ground-based CMB experiment, CMB-S4 \citep{2016arXiv161002743A,2019arXiv190704473A,2019BAAS...51g.209C}, targets a sky patch of less than 10\% sky for an attempt at the detection of primordial gravitational waves.

The observation of limited patches of sky restricts the polarization analysis to a subset only of the sky polarization modes. This does not permit the perfect disambiguation of $E$-type and $B$-type polarization. $E$~modes being significantly brighter than the target $B$ modes (from both gravitational lensing and primordial origin), care must be taken to avoid even a small contribution of $E$ modes in any $B$-mode power spectrum estimate implemented on partial sky observations.

Various methods to address this problem have been proposed by a number of authors. Most popular methods in literature are typically those which construct orthonormal bases for $E$ and $B$ modes, to separate out the `pure' modes from the leakage-causing `ambiguous' modes \cite{Lewis2002, Lewis2003, Bunn2003, SmithZaldarriaga2007, Zhao2010, Kim2010}. There are also methods in pixel space which attempt to clean the leakage from $E$ to $B$ by estimating a template of the leakage in the pixel domain \cite{Liu2019, Liu2019b}. Effective $E$-$B$ mode separation has also been implemented using Wiener filtering methods \cite{Bunn2017,KodiRamanah2018, KodiRamanah2019}. All of these methods vary in complexity and performance. The `pure' mode construction method is however one of the most popular ones for $E$-to-$B$ leakage control. 

In this paper, we investigate a new approach based on the use of ancillary full-sky $E$-mode data to reduce the leakage of $E$ modes in partial sky $B$-mode maps observed with ground-based CMB polarization experiments. The key idea is that outside the small patch of interest, $E$-mode maps with fair signal-to-noise ratio can be used to avoid the sharp discontinuity of the observed $E$-mode polarization map that is the main source of $E$ to $B$ leakage.
As working examples, we consider two simple sky patches that could be observed with a Northern-hemisphere CMB experiment such as the AliCPT telescope currently being deployed on the Ali Observatory site in Tibet \citep{2017arXiv171003047L}, a Southern-hemisphere patch that could be targeted with the future CMB-S4 experiment, and a large sky patch representing observations with the Large Scale Polarization Explorer (LSPE) experiment \citep{LSPE2012}, to study the performance of our method on the large angular scales. We complement them by either a CMB $E$-mode map obtained with the Planck space mission, or by a map obtained with the future LiteBIRD satellite.

The paper is organized as follows: In section \ref{sec:sims} we discuss the simulation setups used in this work. The mathematical framework and background of CMB polarization analysis is discussed in section \ref{sec:analysis}. The ideal case expectation for CMB power spectra estimation for partial sky observations is discussed in \ref{ssec:ideal-case}, a few selected methods of power spectra estimation with $E$ to $B$ leakage reduction are discussed in \ref{ssec:pureB}, and our newly proposed method is discussed in detail in \ref{ssec:method}. We have a comparative discussion of our proposed method in section \ref{sec:discussion}.
\begin{figure}[tbp]
\centering
    \subfloat{\includegraphics[width=0.33\textwidth]{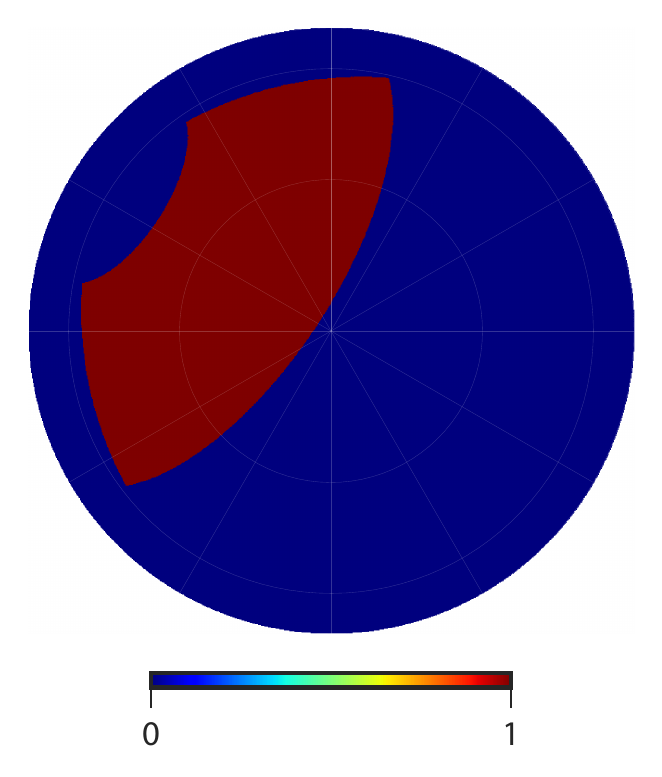}}
    \subfloat{\includegraphics[width=0.33\textwidth]{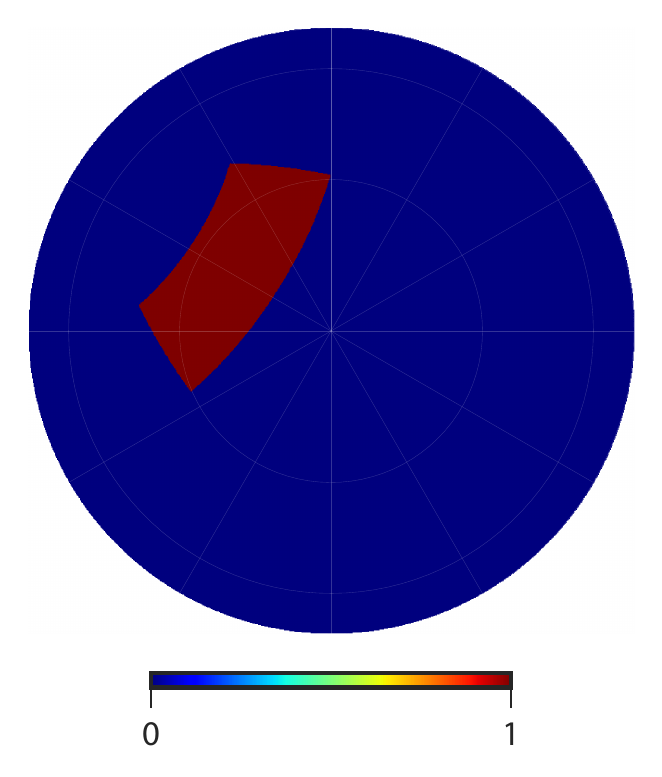}}
    \subfloat{\includegraphics[width=0.33\textwidth]{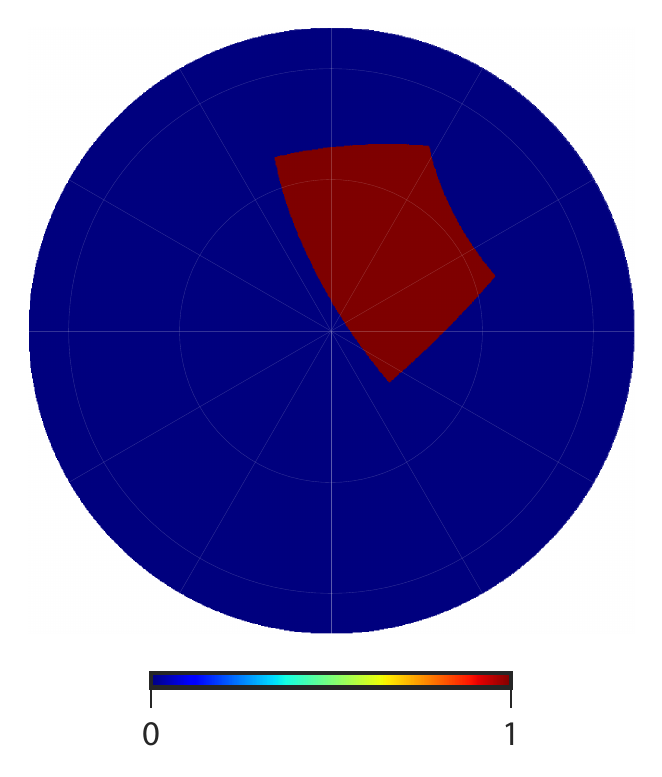}}\\
    \subfloat{\includegraphics[width=0.49\textwidth]{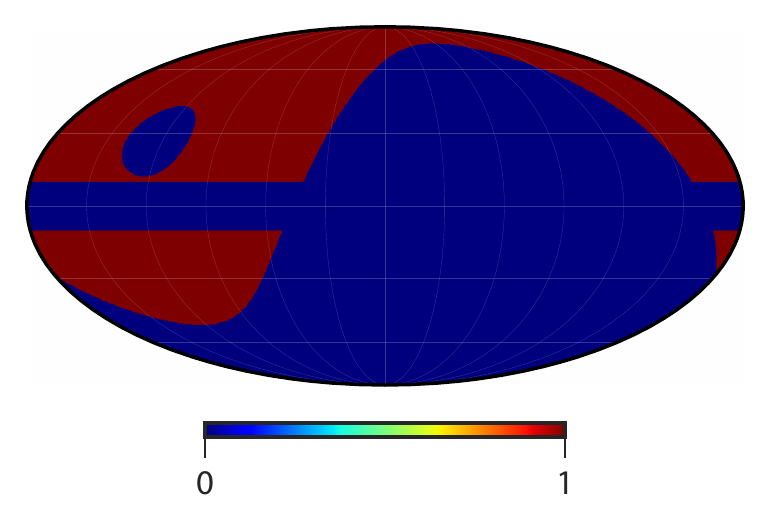}}
    \caption{\label{fig:sky_patches} Orthogonal projection of the three different sky patches studied in this work. Top left: $f_\text{sky}=8\%$ patch 1 in the northern hemisphere, top center: $f_\text{sky}=2\%$ patch 2 in the northern hemisphere, top right: $f_\text{sky}=3\%$ CMB-S4 patch in the southern hemisphere, and bottom: $f_\text{sky}=30\%$ LSPE-SWIPE patch.}
\end{figure} 
\begin{figure}[tbp]
\centering 
\includegraphics[width=0.49\textwidth,trim=160 140 160 100,clip]{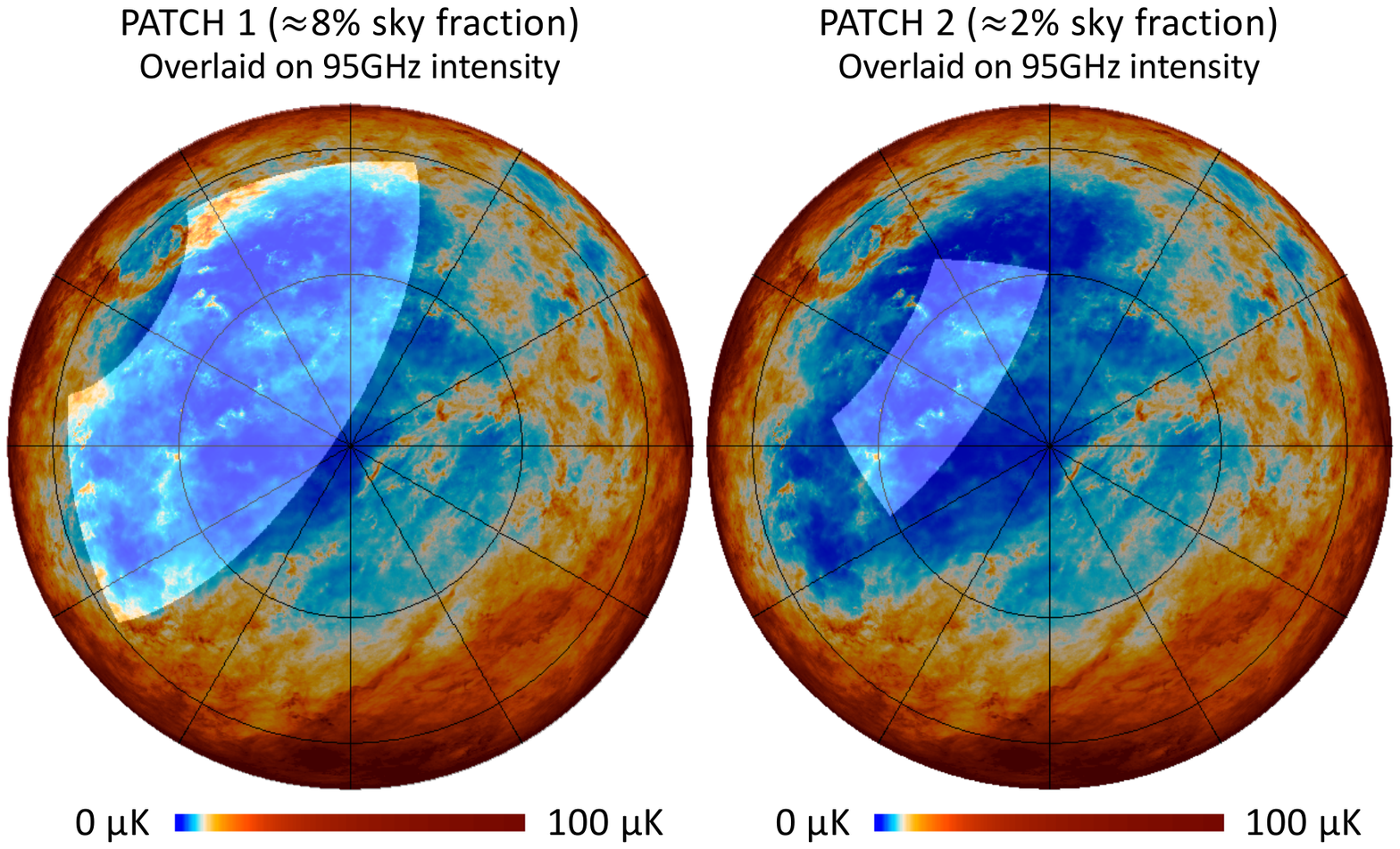}
\hfill
\includegraphics[width=0.49\textwidth,trim=160 140 160 100,clip]{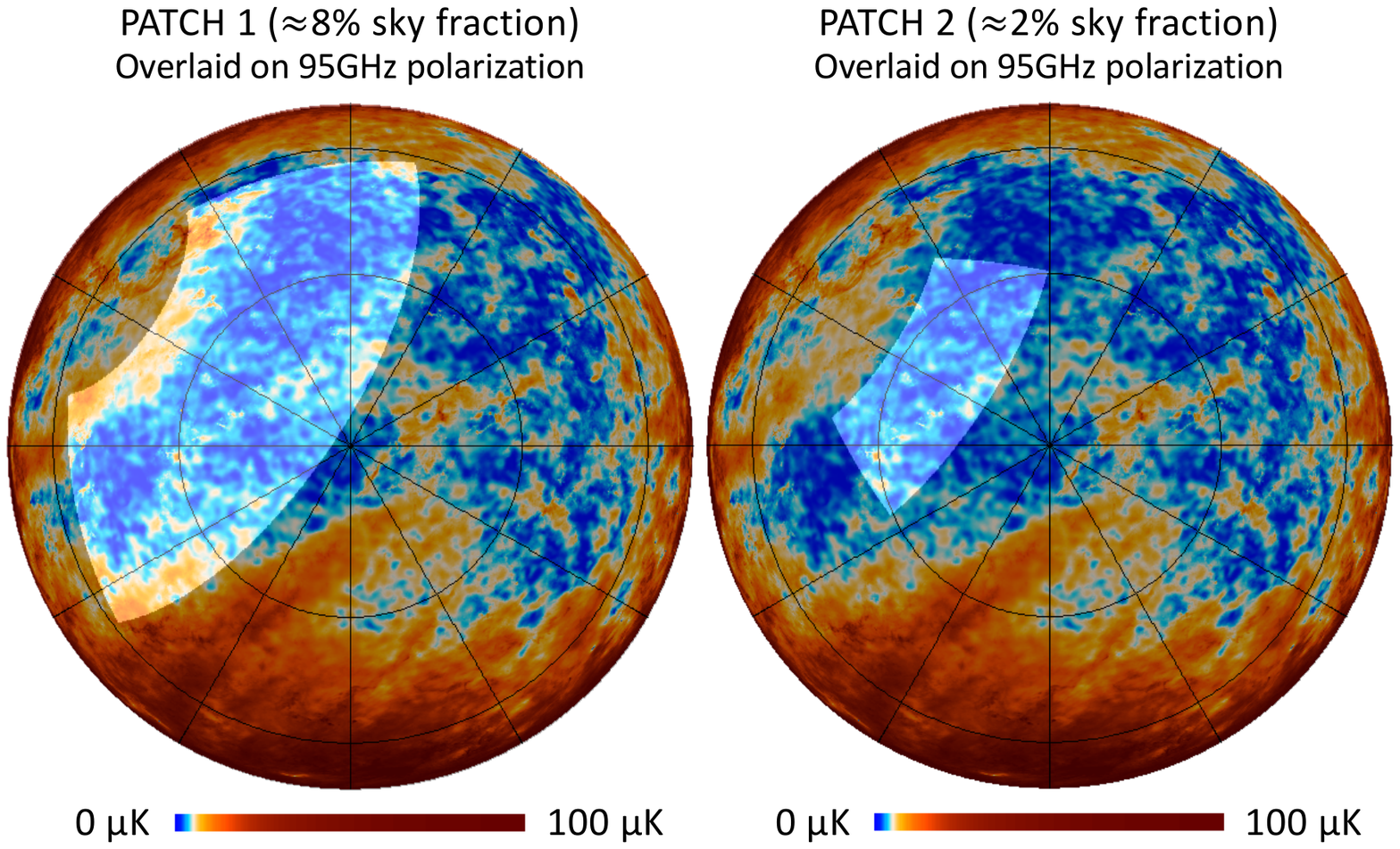}
\hfill
\caption{\label{fig:north-sky-patch} Orthogonal projection of the Northern galactic hemisphere, showing sky patches considered in this work (lighter region). The larger patch (Patch 1, sightly more than 8\% sky fraction), extends from RA=125 to RA=260, and from DEC=30 to DEC=70. The smaller patch extends from RA=158.75 to RA=226.25, and from DEC=40 to DEC=60. The 95\,GHz foreground emission is modelled with an updated version of the Planck Sky Model \citep{2013A&A...553A..96D}. Thermal dust is modelled by scaling a 353\,GHz template from the Planck collaboration \cite{2016A&A...596A.109P}, cleaned from cosmic infrared background contamination with the GNILC method \citep{2011MNRAS.418..467R}. Synchrotron is modelled by scaling in frequency the map of \cite{2015MNRAS.451.4311R}. Dust and Synchrotron polarization, which both contribute to the total polarized emission displayed in the right two panels, are based on Planck collaboration maps from the third data release \citep{2018arXiv180706208P}, obtained with the SMICA method \citep{2003MNRAS.346.1089D,2008ISTSP...2..735C}.}
\end{figure}

\section{Simulations} \label{sec:sims}

In this work we simulate CMB observations for a satellite experiment with full sky coverage, together with observations from a sub-orbital (ground-based or balloon borne) CMB experiment covering a smaller fraction of sky. We consider two limiting cases for the scalar-to-tensor ratio $r$: $r=0$ and $r=0.05$. For the ground-based survey we consider two sky patch options in the northern hemisphere, a `shallow' patch with $f_\text{sky} = 8\%$ and a smaller `deeper' patch with $f_\text{sky} = 2 \%$. We also consider a $f_\text{sky}=3\%$ sky patch in the southern hemisphere with properties of the CMB-S4 Lo-Res Ultra Deep field. To study the estimation of the reionization bump at low-$\ell$s, we consider the LSPE experiment sky patch with $f_\text{sky} = 30\%$. For the LSPE study we only consider the $r=0.05$ case for the reionization bump estimation. These sky patches are shown in figure \ref{fig:sky_patches}. The foreground emissions for the northern hemisphere sky patches are shown in figure \ref{fig:north-sky-patch}. For the CMB-S4 sky patch we choose a 3\% sky patch similar to ones chosen in \cite{CMB-S42019}. The LSPE patch is based on the sky coverage of the balloon borne Short-Wavelength Instrument for the Polarization Explorer (SWIPE) \cite{LSPE2020}, with additional galactic plane masking. We consider full-sky $E$-mode maps obtained with surveys resembling those from the Planck experiment and from the upcoming LiteBIRD space mission. 

We can write the simulated observations $\boldsymbol{d}_X$ as:
\begin{equation}
    \boldsymbol{d}_X = \left(\boldsymbol{s}_X + \boldsymbol{n}_X\right) \cdot \boldsymbol{W}_X,
\end{equation}
where $\boldsymbol{s}_X$ represents the beam-smoothed CMB signal, and $\boldsymbol{n}_X$ the noise realization for instrument $X$. The mask for the corresponding survey region is denoted as $\boldsymbol{W}_X$. In this work we assume negligible foreground contamination residuals, homogeneous noise, and disregard any filtering of the observation timelines to remove systematic effects from ground-pickup and from fluctuations of atmospheric foreground emission. We  postpone our investigation of the impact of these complications to future work. 

\begin{figure}[tbp]
    \centering
    \subfloat{\includegraphics[width=0.5\textwidth]{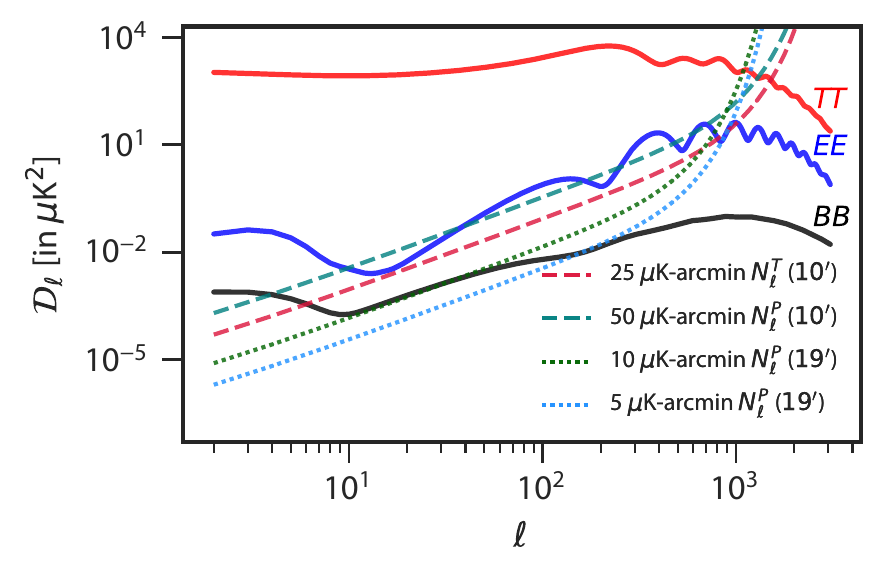}}
    \subfloat{\includegraphics[width=0.5\textwidth]{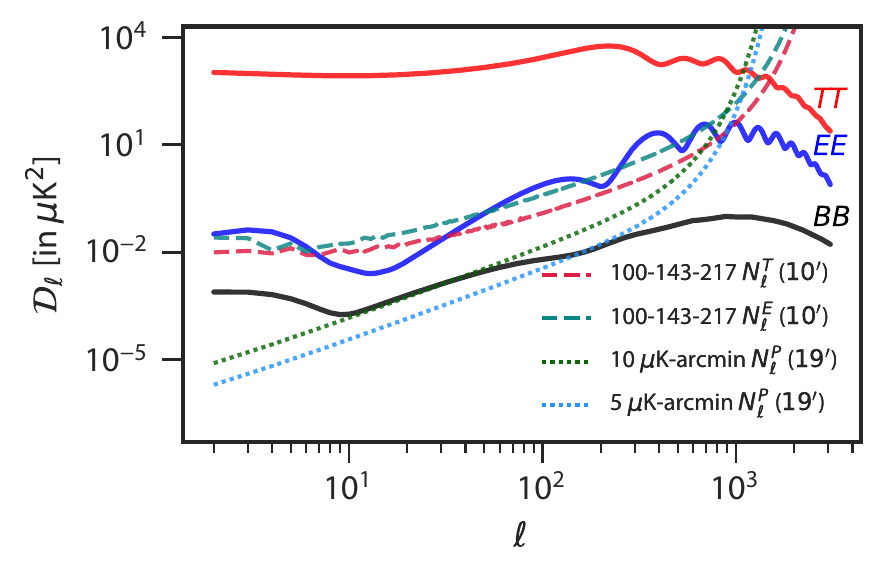}}\\
    \subfloat{\includegraphics[width=0.5\textwidth]{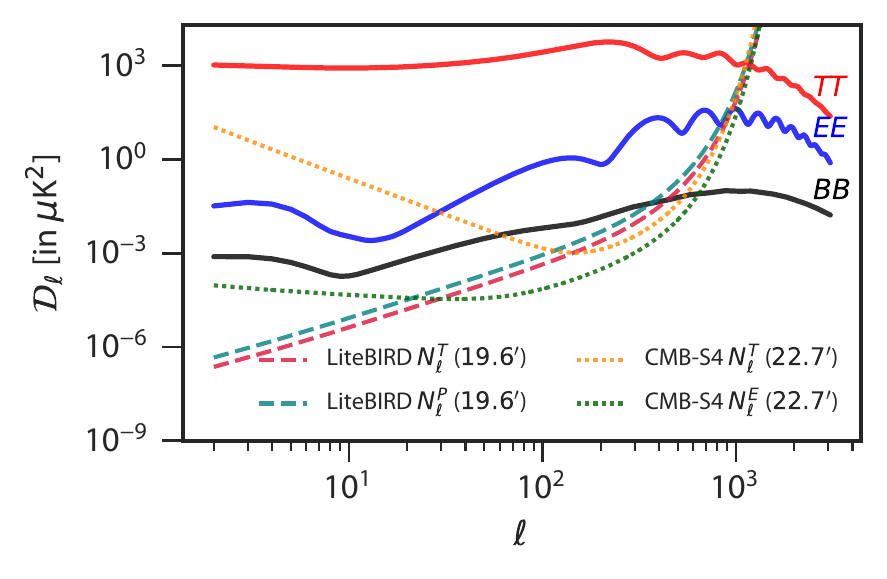}}
    \subfloat{\includegraphics[width=0.5\textwidth]{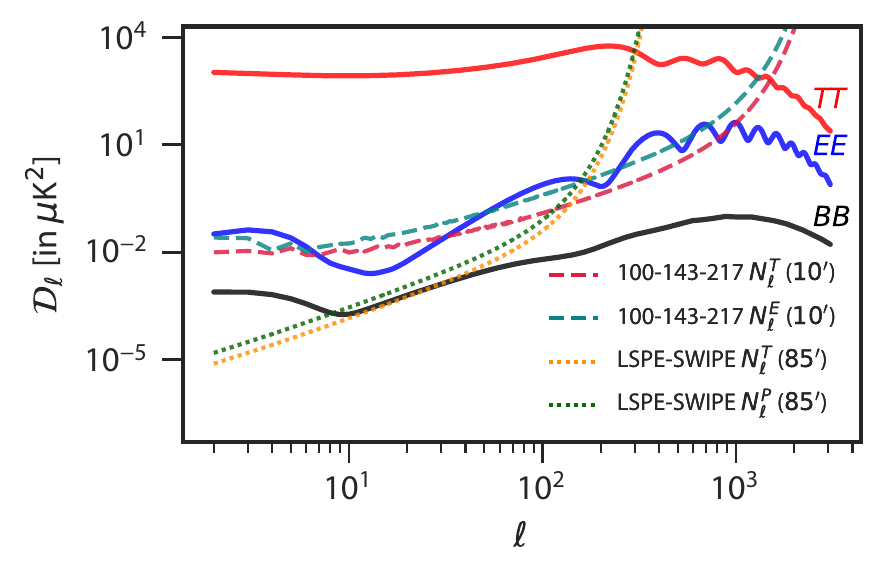}}\\
    \caption{\label{fig:noise_levels} Plot of the noise levels used in this work and comparison with CMB power spectra we use as input for the simulations. Top left: Plot of the noise levels for Planck RMS white-noise levels and patch 1 and patch 2 polarization noise; Top right: Plot of the isotropized FFP10 noise levels for Planck 100-143-217 GHz combined and patch 1 and patch 2 polarization noise; Bottom left: Plot of the noise levels for LiteBIRD and CMB-S4 Lo-Res Ultra-Deep patch; Bottom right: Plot of the isotropized FFP10 noise levels for Planck 100-143-217 GHz combined and LSPE-SWIPE 145-210 GHz combined. The noise power spectra shown in the plots are deconvolved from the beam of the instrument (FWHM mentioned in brackets).}
\end{figure} 
We use the \texttt{synfast} function from the HEALPix\footnote{http://healpix.sourceforge.net} python package (\texttt{healpy}) \cite{Gorski2005, Zonca2019} to generate Gaussian CMB realizations at \texttt{NSIDE}$=1024$. The input CMB power spectra used in \texttt{synfast} are calculated using CAMB\footnote{http://camb.info} \cite{Lewis:1999} with 2018 Planck cosmological parameters \cite{2018arXiv180706209P}, with weak lensing. The northern hemisphere patches are assumed to be observed with an AliCPT-like $19'$ Gaussian beam. The southern hemisphere ground patch is observed with a $22.7'$ Gaussian beam. For LSPE, we assume the sky is observed with the $85'$ Gaussian beam of the SWIPE balloon-borne instrument. The full sky Planck-like simulations are smoothed with a $10'$ Gaussian beam, while the LiteBIRD simulations are smoothed with a $19.6'$ Gaussian beam. This gives us the smoothed CMB signal $\boldsymbol{s}_X$ for the different instruments.

We only consider isotropic white noise in this work. We assume that the northern hemisphere ground-based experiment has polarization white noise RMS of 10 $\mu$K-arcmin in the larger, `shallow' patch (8\% sky fraction) and 5 $\mu$K-arcmin in the smaller, `deeper' patch (2\% sky fraction). For the CMB-S4-like patch in the southern hemisphere (3\% sky fraction) we assume an effective noise at the level expected for the CMB-S4 Small Aperture Telescope (SAT) 95, 145 and 155 GHz channels \citep{CMB-S42019} combined by inverse noise weighting in harmonic space. We compute the noise power spectra for the three channels using values from CMB-S4 Wiki\footnote{https://cmb-s4.org/wiki/index.php/Expected\textunderscore Survey\textunderscore Performance\textunderscore for\textunderscore Science\textunderscore Forecasting} following forecast methods described in \cite{CMB-S42019}, and then adjust for the different resolution of the different channels to bring them all to the common resolution of $22.7'$. The resolution-adjusted noise power spectrum in each frequency band is given by:
\begin{equation}
    \widetilde{N}_{\ell, \nu} = \sigma_P^2\exp{\left[-\frac{\ell(\ell+1)\left(\theta_0^2 - \theta_\nu^2\right)}{8\ln 2}\right]},
    \label{eq:res_adjust}
\end{equation}
where $\sigma_P$ is the RMS of the noise in $\mu$K-radian, $\ell$ denotes the multipole of the spherical harmonic transform, $\theta_\nu$ is the FWHM of the Gaussian beam of the channel centered at frequency $\nu$, and $\theta_0$ the target resolution of $22.7'$ for CMB-S4. We then get an effective noise power spectrum $N_{\ell, \text{eff}}$ for these three channels by combining the noise power spectra ($\tilde N_{\ell, \nu}$) from the three channels,
\begin{equation}
    N_{\ell, \text{eff}} = \left[\sum_\nu \tilde N_{\ell, \nu}^{-1}\right]^{-1}.
    \label{eq:eff_noise}
\end{equation}
We then use \texttt{synfast} to obtain isotropic noise with the CMB-S4 noise power spectra. For the LSPE setup, we consider the baseline noise levels for the SWIPE 145 and 210\,GHz channels. We combine the noise power spectra for the two SWIPE channels at their common resolution of $85'$, using the prescription discussed above for CMB-S4. We generate isotropic noise maps from the combined noise spectra.

For the full sky satellite experiment we consider three noise cases. In the first case, we consider a white noise RMS of 50 $\mu$K-arcmin in polarization and 25 $\mu$K-arcmin in temperature. These values are representative of the sensitivity of the Planck space mission on small and intermediate scales. For our second case, we consider the effective noise power spectra of Planck 100, 143 and 217 GHz HFI channels. We obtain those by averaging noise power spectra from 300 FFP-10 simulated noise maps for each of the three channels. We then use equation \eqref{eq:eff_noise} to obtain the inverse noise weighted effective noise power spectrum of the three combined channels, which matches the value of 50~$\mu$K-arcmin on small scales, but features excess noise for $\ell \leq 100$ (see figure~\ref{fig:noise_levels}, left panel). We use the effective noise power spectra to simulate isotropic noise realizations for the Planck case. 

For the LiteBIRD setup, we combine six Low Frequency Telescope (LFT) channels from 68 to 140 GHz and six High Frequency Telescope (HFT) channels from 100 to 235 GHz \citep{LiteBIRD2018}. These are adjusted for resolution and brought to the common resolution of $19.6'$ following equation \eqref{eq:res_adjust}. Finally these are combined by inverse noise weights by equation \eqref{eq:eff_noise}. A comparison of the input CMB power spectra and the different noise levels is shown in figure~\ref{fig:noise_levels} (right panel). 

Combining the smoothed signal maps and noise maps in the observation patch of the experiment, we get the following set of simulated ground observations:
\begin{enumerate}[topsep=4pt,itemsep=1pt,partopsep=1pt,parsep=1pt]
    \item $\boldsymbol{d}_\text{patch1}$: $19'$ resolution signal with 10 $\mu$K-arcmin noise in patch 1, or
    \item $\boldsymbol{d}_\text{patch2}$: $19'$ resolution signal with 5 $\mu$K-arcmin noise in patch 2; and
    \item $\boldsymbol{d}_\text{CMB-S4}$: $22.7'$ resolution signal with isotropic noise from effective noise power spectra of 95-145-155 GHz channels, in CMB-S4 patch.
    \item $\boldsymbol{d}_\text{LSPE}$: $85'$ resolution signal with isotropic noise from effective noise power spectra of 145-210 GHz channels, in the LSPE-SWIPE patch. This patch is only considered for the case study of the reionization bump.
\end{enumerate}
Similarly, we have the following combined satellite experiment simulations:
\begin{enumerate}[topsep=4pt,itemsep=1pt,partopsep=1pt,parsep=1pt]
    \item $\boldsymbol{d}_\text{Planck1}$: $10'$ resolution signal with 25 $\mu$K-arcmin temperature noise and 50 $\mu$K-arcmin polarization noise, or
    \item $\boldsymbol{d}_\text{Planck2}$: $10'$ resolution signal with isotropic noise from effective noise spectra of the 100-143-217 GHz channels;
    \item $\boldsymbol{d}_\text{LiteBIRD}$: $19.6'$ resolution signal with isotropic noise from the effective noise power spectra from the 12 channels between 68 and 235 GHz.
\end{enumerate}
In the next sections we analyze the performance of $E$-to-$B$ leakage reduction techniques for the three ground observation cases using the simulated data sets.

\section{Partial sky polarization analysis} \label{sec:analysis}
CMB polarization experiments measure Stokes $Q$ and $U$ linear polarization fields on the sky. The Stokes $Q$ and $U$, however, are not scalar fields, as their value depends on the choice of a coordinate system. The combinations $P_{\pm} = Q \pm i U$ are spin-($\pm2$) fields, which can be expanded in terms of spin-weighted spherical harmonics as:
\begin{equation}
    P_{\pm}(\hat n) = \sum_{\ell m} a_{\pm2, \ell m} \, {}_{\pm2}Y_{\ell m}(\hat n),
\end{equation}
where $a_{\pm2, \ell m}$ are the expansion coefficients. We define the scalar $E$-mode  and the pseudoscalar $B$-mode fields with spherical harmonic coefficients \cite{Zaldarriaga1997}: 
\begin{align}
    a_{E, \ell m} &= -\frac{1}{2}\left(a_{2, \ell m} + a_{-2, \ell m}\right) = -\frac{1}{2}\left[\int P_+(\hat n){}_2Y^*_{\ell m}(\hat n)d\Omega +\int P_-(\hat n){}_{-2}Y^*_{\ell m}(\hat n)d\Omega\right] \nonumber\\
    a_{B, \ell m} &= \frac{i}{2}\left(a_{2, \ell m} - a_{-2, \ell m}\right) =\frac{i}{2}\left[\int P_+(\hat n){}_2Y^*_{\ell m}(\hat n)d\Omega - \int P_-(\hat n){}_{-2}Y^*_{\ell m}(\hat n)d\Omega\right].
    \label{eq:EB_sh_coeff}
\end{align}
The $E$- and $B$-mode auto power spectra are defined as:
\begin{align}
    C_\ell^{EE} &= \frac{1}{2\ell + 1} \sum_m \langle a_{E,\ell m} a^*_{E,\ell m}\rangle, \nonumber\\
    C_\ell^{BB} &= \frac{1}{2\ell + 1} \sum_m \langle a_{B,\ell m} a^*_{B,\ell m} \rangle.
\end{align}

One can use spin lowering and raising operators ($\eth$ and $\bar \eth$) of \cite{Newman1966} to construct spin zero fields from the spin-($\pm$2) fields. We can define two rotationally invariant fields in real space as \cite{Zaldarriaga1997}:
\begin{align}
    \mathcal{E}(\hat n) &= -\frac{1}{2}\left[ \bar \eth \bar \eth P_+(\hat n) + \eth \eth P_-(\hat n)\right], \nonumber \\
    \mathcal{B}(\hat n) &= \frac{i}{2}\left[ \bar \eth \bar \eth P_+(\hat n) - \eth \eth P_-(\hat n)\right].
    \label{eq:pure_EB}
\end{align}
The $\mathcal{E}$ and $\mathcal{B}$ fields are the pure-$E$ and the pure-$B$ fields. These can be related to potential functions $\mathcal{E} = \eth \eth \bar \eth \bar \eth \psi_\mathcal{E}$, and $\mathcal{B} = \eth \eth \bar \eth \bar \eth \psi_\mathcal{B}$ \cite{Bunn2003}. So we can write $P_+ = - \eth \eth \left[\psi_\mathcal{E} + i \psi_\mathcal{B}\right]$ and $P_- = - \bar \eth \bar \eth \left[\psi_\mathcal{E} - i \psi_\mathcal{B}\right]$. It can be seen that a pure-$E$ only polarization field has no pure-$B$ projection or vise-versa. Hence the pure-$E$ and pure-$B$ fields are orthogonal to one another. The pure $E$-$B$ fields are related to the $E$-$B$ modes by:
\begin{equation}
    a_{\mathcal{E}/\mathcal{B}, \ell m} = \sqrt{\frac{(\ell +2)!}{(\ell-2)!}}a_{E/B, \ell m}.
\end{equation}

For a partial sky analysis, we have two complications. First, because of the incomplete sky coverage, there is mixing of power between the different harmonic modes. Even for a scalar field, like the CMB temperature field, we would need to correct for mode mixing. The most common method to correct for this is by the so-called pseudo-$C_\ell$ estimators \citep{Hivon2002}. The second additional complication arises for CMB polarization as the decomposition into $E$ and $B$ modes is unique only for full sky $Q$ and $U$ fields. For partial sky $Q$ and $U$ observations, modes which satisfy the properties of either $E$ or $B$ modes project similarly on the partial sky patch. These are termed `ambiguous' modes \cite{Bunn2003}. The contribution of these ambiguous modes to the $B$-mode  spectrum generates a significant overestimate of the $B$-mode  power spectrum, termed as $E$-to-$B$ leakage. 

On a part of sky defined by the weight function $W(\hat n)$, the $E$- and $B$-mode spherical harmonic coefficients are given by:
\begin{align}
    \tilde a_{E, \ell m} &= -\frac{1}{2}\left[\int P_+(\hat n)W(\hat n){}_2Y^*_{\ell m}(\hat n)d\Omega +\int P_-(\hat n)W(\hat n){}_{-2}Y^*_{\ell m}(\hat n)d\Omega\right], \nonumber\\
    \tilde a_{B, \ell m} &= \frac{i}{2}\left[\int P_+(\hat n)W(\hat n){}_2Y^*_{\ell m}(\hat n)d\Omega - \int P_-(\hat n)W(\hat n){}_{-2}Y^*_{\ell m}(\hat n)d\Omega\right].
\end{align}
We can rewrite the full sky $P_\pm$ field, and the window function in terms of their spherical harmonic decomposition in the above relations giving us:
\begin{align}
    \tilde a_{E, \ell m} &= \sum_{\ell' m'}\left[ K_{\ell m \ell' m'}^{EE}a_{E, \ell' m'} + iK_{\ell m \ell' m'}^{EB}a_{B, \ell' m'}\right], \nonumber\\
    \tilde a_{B, \ell m} &= \sum_{\ell' m'}\left[ -iK_{\ell m \ell' m'}^{BE}a_{E, \ell' m'} + K_{\ell m \ell' m'}^{BB}a_{B, \ell' m'}\right].
    \label{eq:EB_partsky}
\end{align}
The mixing kernels $K^{XY}_{\ell m \ell' m'}$ \cite{Ferte2013} mix the $E$ and $B$ modes, as well as spherical harmonic modes corresponding to different $(\ell,m)$ pairs. It should be noted that the $K^{EB}_{\ell m \ell' m'}$ matrix couples the partial sky $E$ modes ($\tilde a_{E, \ell m}$) with the original $B$ modes ($a_{B, \ell m}$), while the $K^{BE}_{\ell m \ell' m'}$ matrix couples the partial sky $B$ modes ($\tilde a_{B, \ell m}$) with the original $E$ modes ($a_{E, \ell m}$). These are the leakage terms for the partial sky decomposition.

In the following, we first discuss the power spectra recovery from partial sky data, then outline some of the existing methods to correct for the $E$-to-$B$ leakage, and finally introduce our new method for $E$-to-$B$ leakage control.  

\subsection{The ideal case performance}  \label{ssec:ideal-case}
For a scalar field $f(\hat n) = \sum_{\ell m} a_{\ell m}Y_{\ell m}(\hat n)$, spherical harmonic coefficients on a partial sky $\tilde a_{\ell m}$ can be written as:
\begin{align}
    \tilde a_{\ell m} &= \int f(\hat n) W(\hat n) Y^*_{\ell m}(\hat n) d \Omega \\
    &= \sum_{\ell' m'} K_{\ell m \ell' m'} a_{\ell' m'},
\end{align}
where $W(\hat n)$ defines the patch of sky `seen' by the observation and $K_{\ell m \ell' m'}$ is the mixing kernel \cite{Hivon2002}. This shows the mixing of harmonic modes due to partial sky observation. At power spectrum level we have:
\begin{equation}
    \langle \tilde C_\ell \rangle = \sum_{\ell' \ell''} \frac{(2\ell' + 1) (2\ell'' + 1) }{4 \pi} W_{\ell''} \begin{pmatrix}
\ell & \ell' & \ell''\\
0 & 0 & 0
\end{pmatrix}^2 \langle C_{\ell'}\rangle = \sum_{\ell'} M_{\ell \ell'}  \langle C_{\ell'}\rangle,
\label{eq:scalar_mix}
\end{equation}
where $W_{\ell}$ is the power spectrum of the weight function $W(\hat n)$, and $M_{\ell \ell'}$ is called the mixing matrix. 

Since the $E$ and $B$ modes define a scalar and a pseudoscalar field, we can use the $E$-mode and $B$-mode spherical harmonic coefficient of equation \eqref{eq:EB_sh_coeff} to define two scalar fields:
\begin{align}
    E(\hat n) &= \sum_{\ell m} a_{E, \ell m} Y_{\ell m}(\hat n), \nonumber\\
    B(\hat n) &= \sum_{\ell m} a_{B, \ell m} Y_{\ell m}(\hat n).
\end{align}
In the case where there is no leakage from $E$ to $B$, we can treat the $E$ or $B$ mode as a decoupled scalar field. Then for partial sky analysis of $E$ or $B$ fields without leakage we can use equation \eqref{eq:scalar_mix} to correct for the mixing of power between different harmonic modes (mode mixing). For this ideal situation the power spectrum estimate is unbiased and the error is minimal. It is common practice to bin the estimated power spectra in multipole bins to average over random fluctuations. The  binned spectrum  $\mathcal{D}_{\bar \ell}$, in a bin of size $\Delta \ell$, centered about $\bar \ell$ is then given as:
\begin{equation}
    \mathcal{D}_{\bar \ell} = \sum_{\ell = \bar \ell -\Delta \ell/2}^{\ell = \bar \ell +\Delta \ell/2} \frac{\ell \left(\ell + 1\right)}{2 \pi \Delta \ell} C_\ell.
\end{equation}
For observations from ground-based experiments dedicated to the detection of primordial $B$-modes, we restrict ourselves to the multipole range of $20 \leq \ell \leq 500$. For partial sky analysis on a patch with sky fraction $f_\text{sky}$, observed with an axisymmetric beam with Legendre coefficients $B_\ell$, with a binned noise power spectrum $\mathcal{N}_{\bar \ell}$, the optimal error is given as \cite{Hivon2002,Challinor2005}: 
\begin{equation}
    \Delta \mathcal{D}_{\bar \ell; \text{optimal}} = \sqrt{\frac{2}{(2\bar \ell + 1) f_\text{sky} \Delta \ell}} \left[ \mathcal{D}_{\bar \ell} + \frac{\mathcal{N}_{\bar \ell}}{B^2_{\bar \ell}} \right] \left[\frac{w_{(4)}}{w_{(2)}^2}\right]^{1/2}. 
    \label{eq:cosmic_var}
\end{equation}
Here $B^2_{\bar \ell}$ is the mean squared beam value in the multipole bin, while $4 \pi f_\text{sky} w_{(i)} = \int |W(\hat n)|^i d\Omega$. The $w_{(4)}/w_{(2)}^2$ is the correction factor due to apodization \cite{HGH2002}. In this work we are looking for a method of partial sky polarization analysis that will give us an unbiased estimate of the $B$-mode  power spectrum, free from $E$-to-$B$ leakage, and with errors equal to the ideal case discussed above. 

\subsection{Partial sky analysis with E-B leakage control} \label{ssec:pureB}
In the usual partial sky analysis we know that there is $E$-to-$B$ leakage due to coupling term $K^{BE}_{\ell m \ell' m'}$. The power spectra for $E$ and $B$ modes are given by:
\begin{align}
    \langle \tilde C_\ell^{EE}\rangle &= \sum_{\ell'} \left[ M^{EE}_{\ell \ell'} \langle C_{\ell'}^{EE} \rangle + M^{EB}_{\ell \ell'} \langle C_{\ell'}^{BB} \rangle \right], \nonumber\\
    \langle \tilde C_\ell^{BB}\rangle &= \sum_{\ell'} \left[ M^{BE}_{\ell \ell'} \langle C_{\ell'}^{EE} \rangle + M^{BB}_{\ell \ell'} \langle C_{\ell'}^{BB} \rangle \right],
    \label{eq:pol_EB_mix}
\end{align}
where $M^{XY}_{\ell \ell'}$ is the mixing matrix for CMB polarization, which are calculated from the mixing kernels as:
\begin{equation}
    M^{XY}_{\ell \ell'} = \sum_{m m'} \frac{1}{2\ell + 1}|K^{XY}_{\ell m \ell' m'}|^2.
\end{equation}

Detailed forms of the mixing matrix for polarization can be found in \cite{Alonso2019}. This is the standard pseudo-$C_\ell$ (PCL) method for CMB polarization power spectra estimation. It accounts for the $E$-to-$B$ leakage via the mixing matrix. By inverting the equations \eqref{eq:pol_EB_mix} one can correct for mode mixing and $E$-to-$B$ leakage. However, this is done with a noise penalty due to the inversion of the linear system, so that the final errors can be substantially larger than the ideal lower bound. To test the performance of the standard PCL method in $E$-to-$B$ leakage control we have implemented it with the python package of NaMaster\footnote{https://github.com/LSSTDESC/NaMaster} \cite{Alonso2019}. We have used C2 apodization \cite{Grain2009} for the spherical harmonic transformations for all power spectrum estimation. Throughout this work, we have suitably debiased the power spectra estimates with the average noise power spectra obtained from 400 noise-only simulations. We compute the power spectrum estimates for standard PCL for the following cases: I. $\boldsymbol{d}_\text{patch1}$ for the `shallow', larger patch and II. $\boldsymbol{d}_\text{patch2}$ for the `deep', smaller patch.  The results for cases I and II are shown in figures \ref{fig:fsky8pc_50-10} and \ref{fig:fsky2pc_50-5} respectively. We have plotted the mean power spectrum estimate from 300 simulations with the error bars equal to the standard deviation. We see from those plots that the performance of the standard PCL method is not optimal, in particular in the 2\% sky patch case for which the error bars for $\ell \leq 100$ are very large as compared to the theoretical optimum. 

The most common method used to tackle the $E$-to-$B$ leakage problem is to ignore the ambiguous modes, which are those that contribute most to the total uncertainty. To do this, one works with the pure-$B$ field of equation \eqref{eq:pure_EB}. Then the partial sky $B$ modes can be calculated from the pure-$B$ field, which is constructed to be orthogonal to all the $E$ modes (pure or ambiguous):
\begin{equation}
    \tilde a_{B, \ell m} = \sqrt{\frac{(\ell -2)!}{(\ell + 2)!}} \int W(\hat n) \mathcal{B}(\hat n) Y^*_{\ell m} d\Omega.
    \label{eq:alm_puremthd}
\end{equation}
There are several different approaches for using pure-$B$ construction for $E$ to $B$ leakage reduction \cite{SmithZaldarriaga2007, Zhao2010, Kim2010}. In this work we use the Smith-Zaldarriaga approach \cite{SmithZaldarriaga2007}, since it is the best performing implementation \cite{Ferte2013}. We then proceed with the pseudo-$C_\ell$ method with the pure-$B$ field. When the partial sky $B$-modes are computed with the pure-$B$ method the mixing matrix $M^{BE}_{\ell \ell'}$ (that controls the $E$-to-$B$ leakage term) becomes few orders smaller than the standard case. This reduces $E$-to-$B$ leakage dramatically. For this work we have implemented the pure-$B$ pseudo-$C_\ell$ with the python package of NaMaster, for all the four suborbital experiments considered here: I. $\boldsymbol{d}_\text{patch1}$, II. $\boldsymbol{d}_\text{patch2}$, III. $\boldsymbol{d}_\text{CMB-S4}$, and IV. $\boldsymbol{d}_\text{LSPE}$. The results for cases I and II are shown in figures \ref{fig:fsky8pc_50-10} and \ref{fig:fsky2pc_50-5} respectively, and results for case III is shown in figure \ref{fig:liteBIRD-CMB-S4}. We show the mean power spectrum estimate from 300 simulations with errors given by the standard deviation. The result from 800 simulations for case IV is plotted in figure \ref{fig:Planck-LSPE}, showing the power estimates for the reionization bump. We see that while pure-$B$ method has optimal error bars at high-$\ell$, the performance becomes sub-optimal at low multipoles. With smaller $f_\text{sky}$  observation patch, the performance of pure-$B$ method deteriorates.

In figures \ref{fig:fsky8pc_FFP10-10} and \ref{fig:fsky2pc_FFP10-5} the results for the pure-$B$ method for patches 1 and 2 are displayed again, and are identical to those shown in figures \ref{fig:fsky8pc_50-10} and \ref{fig:fsky2pc_50-5}. However, they are plotted together with results for different cases of our new method for the benefit of comparison (see next section).

\subsection{Partial sky analysis with ancillary full-sky data} \label{ssec:method}

\begin{figure}[tbp]
\centering 
\subfloat{\includegraphics[width=.33\textwidth]{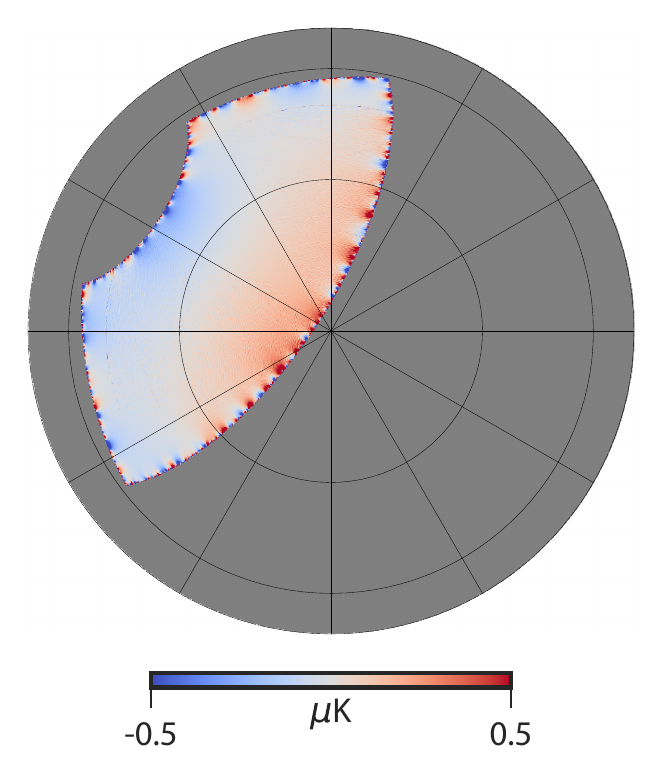}}
\subfloat{\includegraphics[width=.33\textwidth]{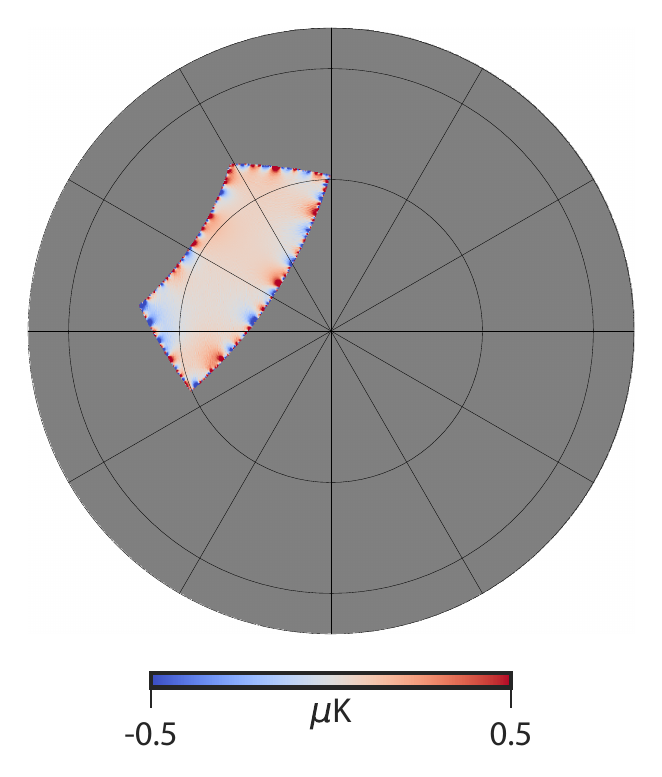}}
\subfloat{\includegraphics[width=.33\textwidth]{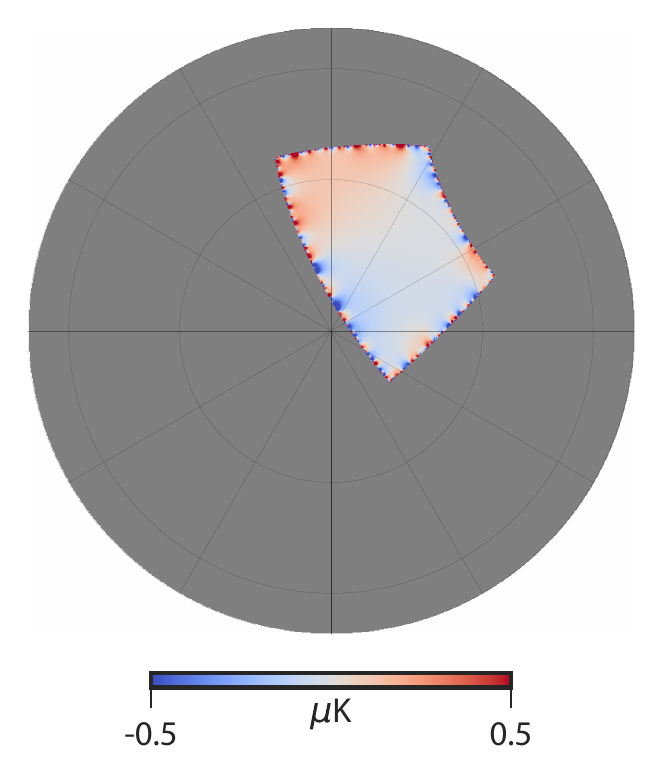}} \\
\subfloat{\includegraphics[width=.49\textwidth]{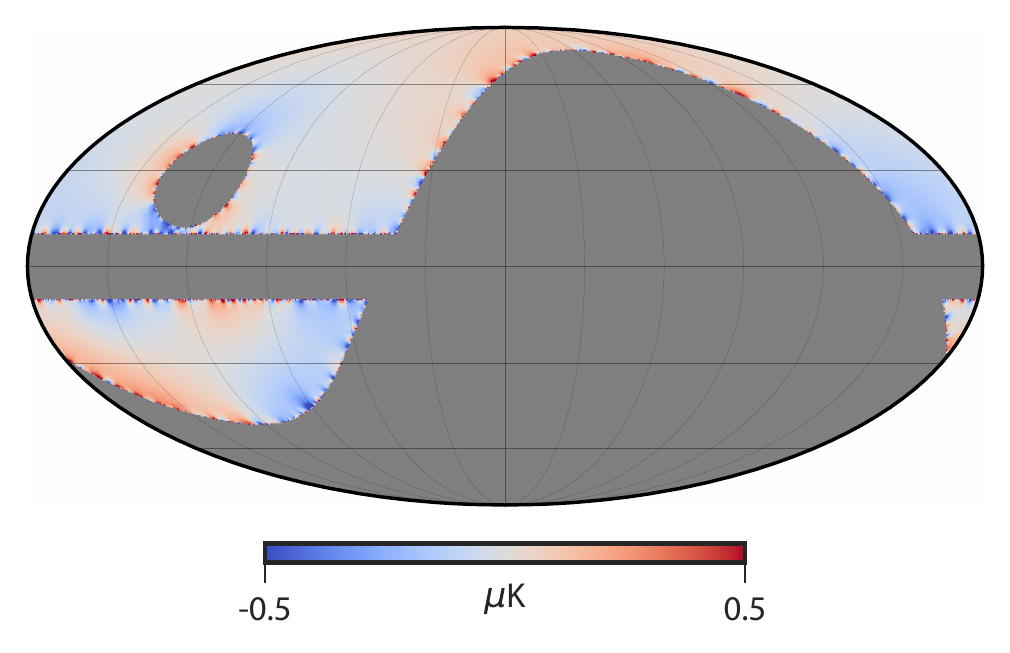}}
\caption{\label{fig:leakage-maps} $E$-to-$B$ leakage for the three sky patches considered here. Top left: north sky patch 1; Top center: north sky patch 2; Top right: south sky CMB-S4 patch; Bottom: LSPE-SWIPE patch.}
\end{figure}

Most of the $E$-to-$B$ leakage in a partial sky harmonic analysis arises from ambiguous modes at the edge of the observed patch. This is easily visualized by generating a CMB map with no $B$ modes (signal or noise), masking the region outside of the observed patch, computing $T$-, $E$- and $B$-mode harmonic coefficients with a spherical harmonic transform of the masked temperature and polarization maps, and performing a back transform of the $B$-mode  harmonic coefficients alone to form a map of the $E$-to-$B$ leakage. For the sky patches considered here as an example, the resulting `leakage maps', restricted to the observed regions, are shown in figure \ref{fig:leakage-maps}. 

\begin{figure}[tbp]
\centering
    \subfloat{\includegraphics[width=0.5\textwidth]{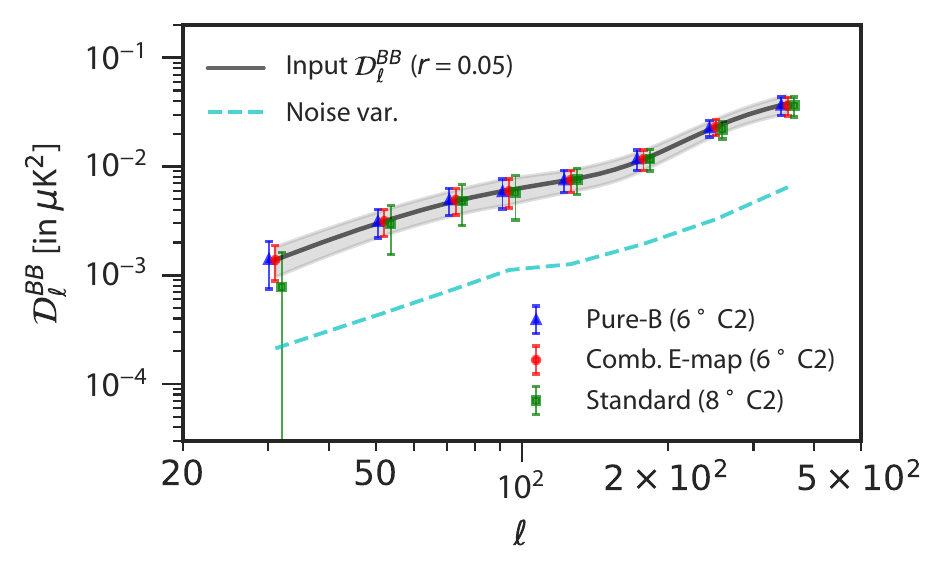}}
    \subfloat{\includegraphics[width=0.5\textwidth]{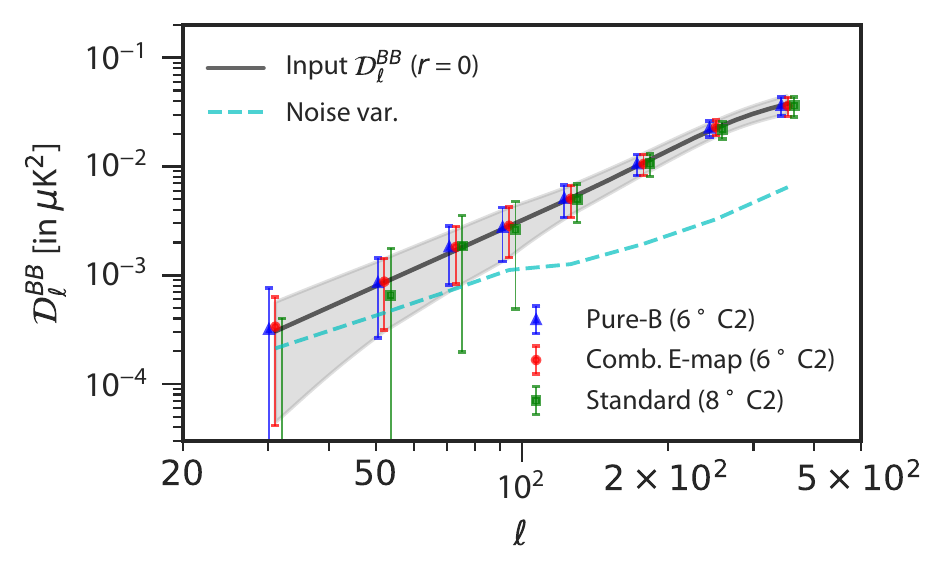}}\\
    \caption{Comparison of $B$-mode power spectrum estimated with different methods for the 8\% $f_\text{sky}$ patch data: $\boldsymbol{d}_\text{patch1}$. The power spectra are estimated by standard polarization PCL method (in green), pure-$B$ PCL method (in blue), and map combination method using filtered Planck-like simulations with 50 $\mu$K-arcmin noise ($\boldsymbol{\widehat d}_\text{Planck1}$), and scalar PCL (our method for case I, in red). The black curve represent the input CMB $B$-mode  power spectrum ($r$ = 0.05 on left and $r$ = 0 on right). The grey region shows the optimal error bounds. The dashed cyan curve shows the noise contribution to the optimal error limits. }
    \label{fig:fsky8pc_50-10}
\end{figure}
\begin{figure}[tbp]
\centering
    \subfloat{\includegraphics[width=0.5\textwidth]{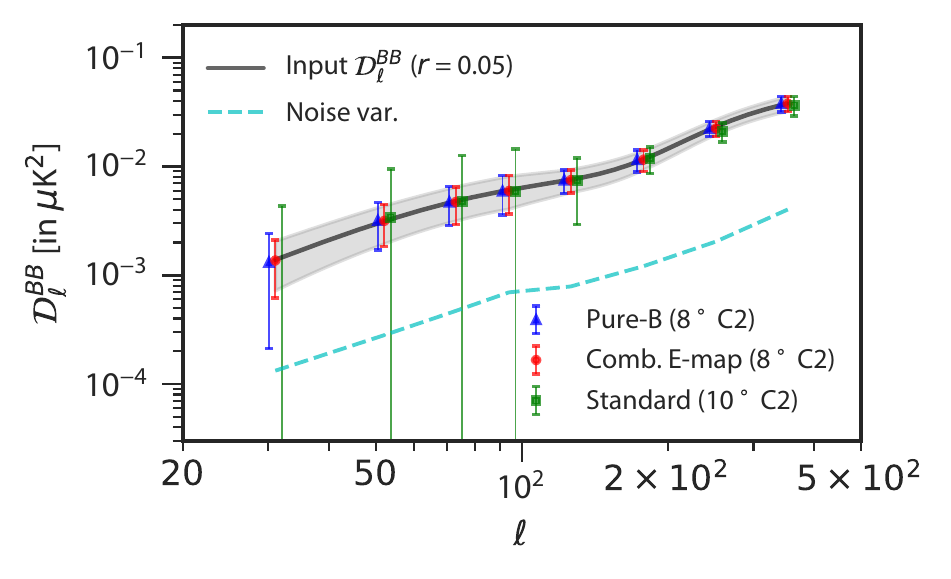}}
    \subfloat{\includegraphics[width=0.5\textwidth]{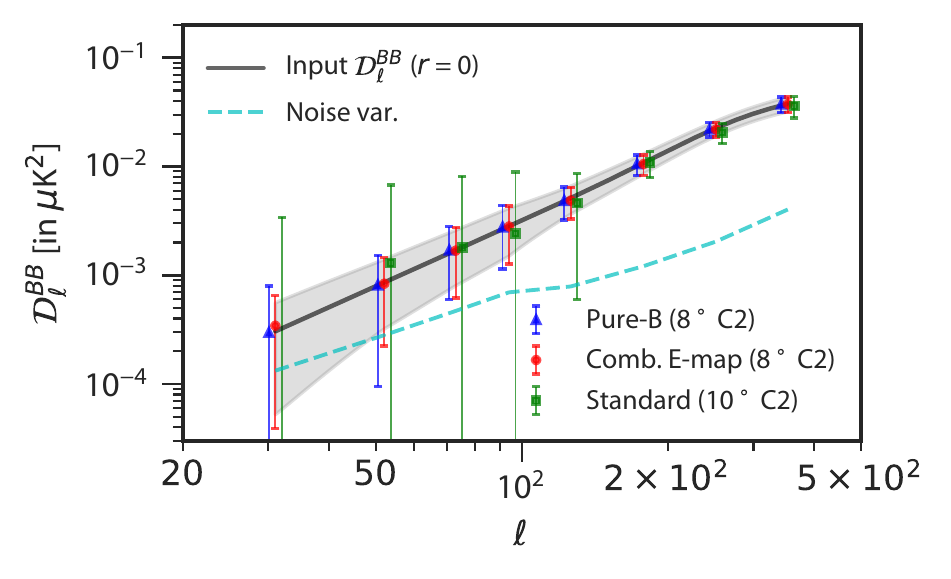}}\\
    \caption{Comparison of $B$-mode power spectrum estimated with different methods for the 2\% $f_\text{sky}$ patch data: $\boldsymbol{d}_\text{patch2}$. The power spectra are estimated by standard polarization PCL method (in green), pure-$B$ PCL method (in blue), and map combination method using filtered Planck-like simulations with 50 $\mu$K-arcmin noise ($\boldsymbol{\widehat d}_\text{Planck1}$), and scalar PCL (our method for case II, in red). The black curve represent the input CMB $B$-mode  power spectrum ($r$ = 0.05 on left and $r$ = 0 on right). The grey region shows the optimal error bounds. The dashed cyan curve shows the noise contribution to the optimal error limits. }
    \label{fig:fsky2pc_50-5}
\end{figure}

We know that the $E$-to-$B$ leakage occurs due to the incompleteness of the $E$-mode signal outside the observation patch. So, to reduce the ambiguous modes arising from the incomplete $E$-mode signal, one can make use of any additional data that provide a way to estimate the $E$ modes outside the observed region, and in particular in pixels at a distance corresponding to the typical correlation length of CMB $E$ modes. One can use CMB space mission data, like that from the Planck space mission, where the $E$ modes are measured with a fair signal-to-noise ratio, to complete the $E$-mode signal outside the observation region.

To implement $E$-to-$B$ leakage reduction using space mission data outside the observed region, we first build a full-sky minimum-variance map of $E$ modes from simulated full sky observations ($\boldsymbol{d}_\text{Planck1}$, $\boldsymbol{d}_\text{Planck2}$ or $\boldsymbol{d}_\text{LiteBIRD}$). To that effect, we use both temperature and $E$-mode polarization observations, and make a map of estimated $E$ modes at the angular resolution of the ground-based experiment being considered. 

\begin{figure}[tbp]
\centering
    \subfloat{\includegraphics[width=0.5\textwidth]{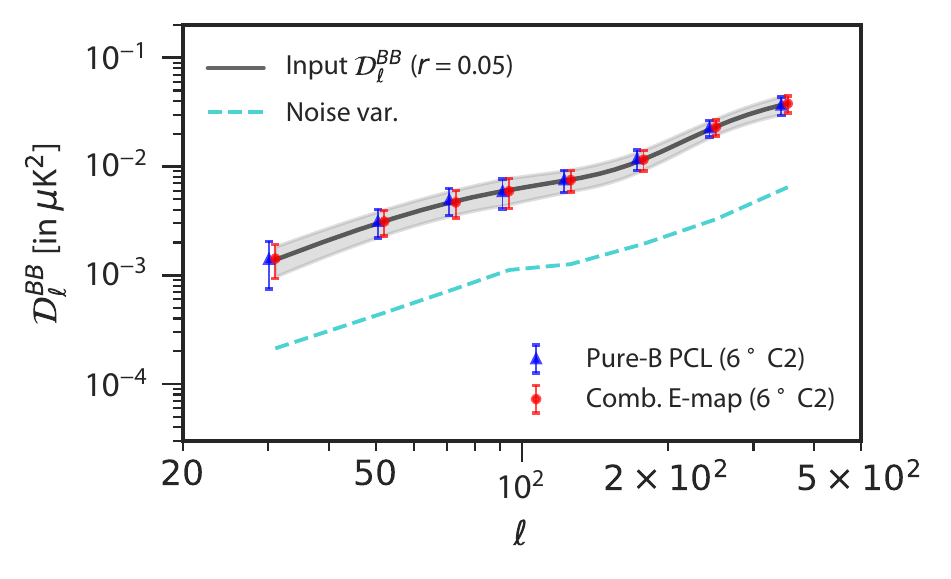}}
    \subfloat{\includegraphics[width=0.5\textwidth]{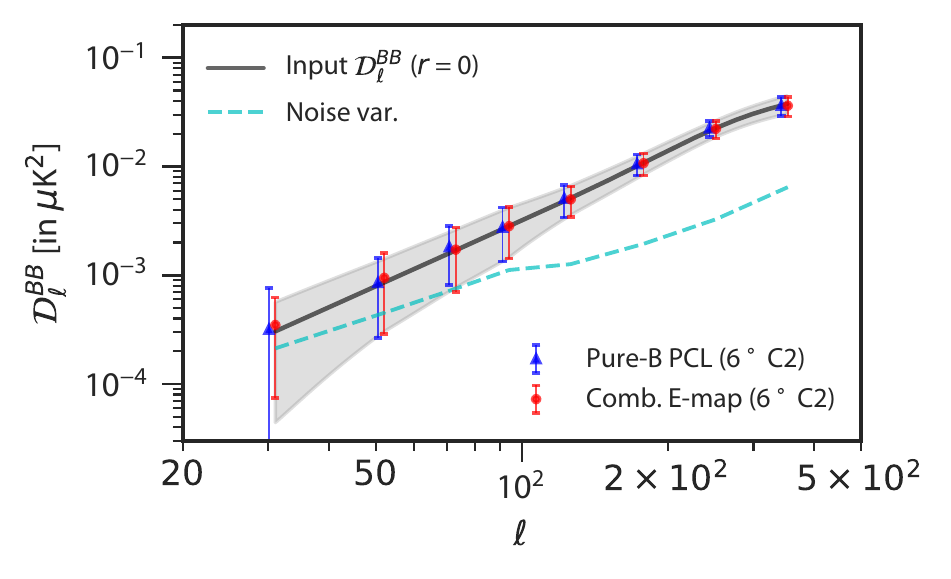}}\\
    \caption{Comparison of $B$-mode  power spectrum estimated with different methods for the 8\% $f_\text{sky}$ patch data: $\boldsymbol{d}_\text{patch1}$. The power spectra are estimated by pure-$B$ PCL method (in blue), and map combination method using filtered Planck-like simulations with FFP10 effective noise level  ($\boldsymbol{\widehat d}_\text{Planck2}$), and scalar PCL (our method for case III, in red). The black curve represent the input CMB $B$-mode power spectrum ($r$ = 0.05 on left and $r$ = 0 on right). The grey region shows the optimal error bounds. The dashed cyan curve shows the noise contribution to the optimal error limits. }
    \label{fig:fsky8pc_FFP10-10}
\end{figure}
\begin{figure}[tbp]
\centering
    \subfloat{\includegraphics[width=0.5\textwidth]{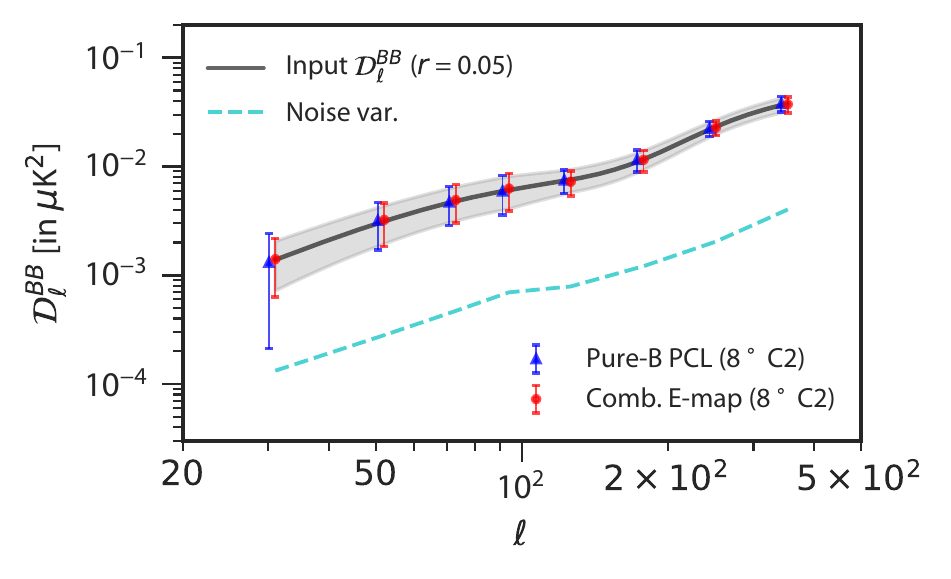}}
    \subfloat{\includegraphics[width=0.5\textwidth]{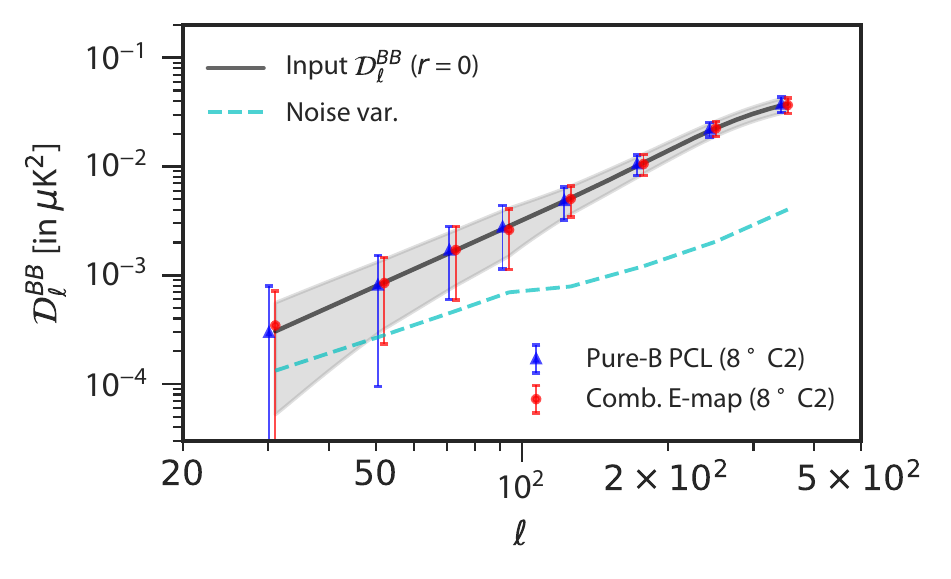}}\\
    \caption{Comparison of $B$-mode  power spectrum estimated with different methods for the 2\% $f_\text{sky}$ patch data: $\boldsymbol{d}_\text{patch2}$. The power spectra are estimated by pure-$B$ PCL method (in blue), and map combination method using filtered Planck-like simulations with FFP10 effective noise level ($\boldsymbol{\widehat d}_\text{Planck2}$), and scalar PCL (our method for case IV, in red). The black curve represent the input CMB $B$-mode power spectrum ($r$ = 0.05 on left and $r$ = 0 on right). The grey region shows the optimal error bounds. The dashed cyan curve shows the noise contribution to the optimal error limits. }
    \label{fig:fsky2pc_FFP10-5}
\end{figure}
\begin{figure}[tbp]
\centering
    \subfloat{\includegraphics[width=0.5\textwidth]{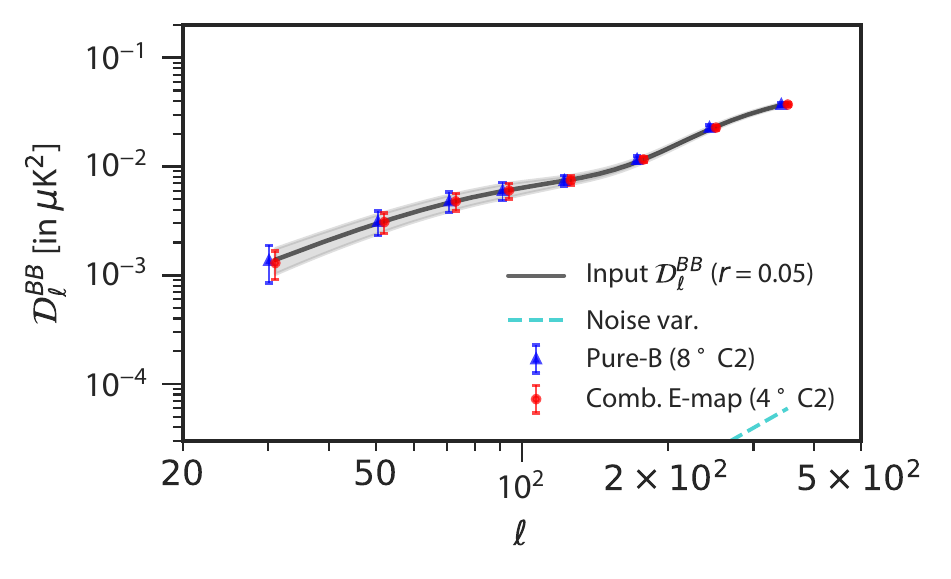}}
    \subfloat{\includegraphics[width=0.5\textwidth]{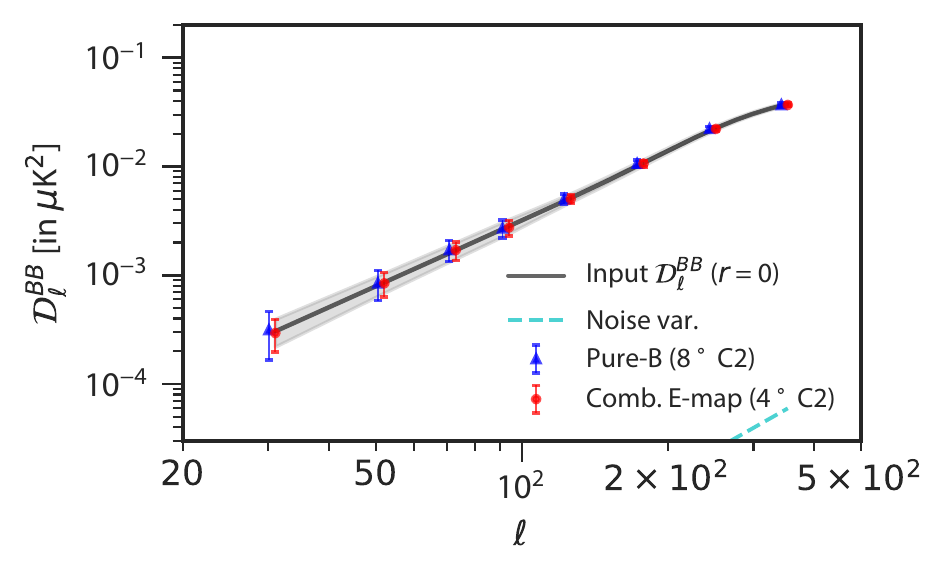}}\\
    \caption{Comparison of $B$-mode  power spectrum estimated with different methods for the CMB-S4 patch data: $\boldsymbol{d}_\text{CMB-S4}$. The power spectra are estimated by pure-$B$ PCL method (in blue), and map combination method using filtered LiteBIRD-like simulations ($\boldsymbol{\widehat d}_\text{LiteBIRD}$), and scalar PCL (our method for case V, in red). The black curve represent the input CMB $B$-mode power spectrum ($r$ = 0.05 on left and $r$ = 0 on right). The grey region shows the optimal error bounds. The dashed cyan curve shows the noise contribution to the optimal error limits. }
    \label{fig:liteBIRD-CMB-S4}
\end{figure}
\begin{figure}[tbp]
\centering
    \includegraphics[width=0.5\textwidth]{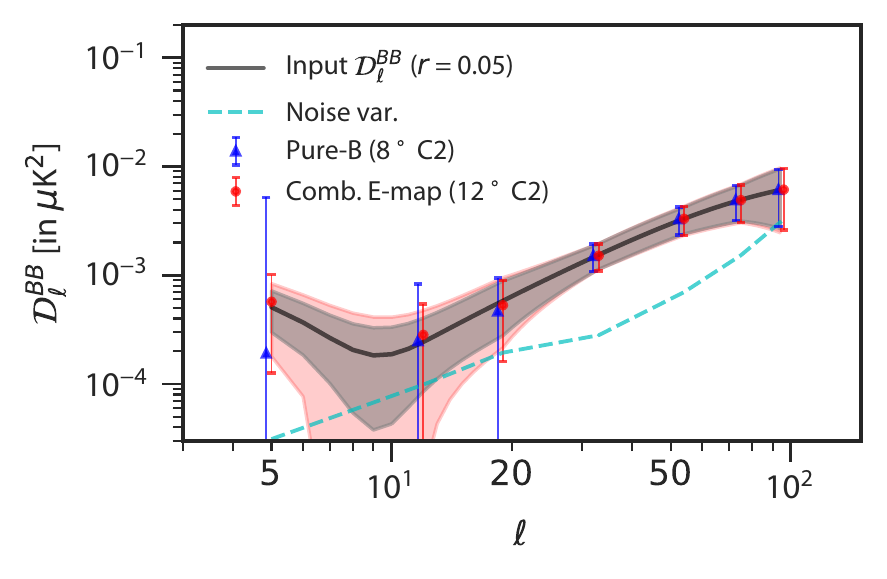}
    \caption{Comparison of $B$-mode  power spectrum estimated with different methods for the LSPE-SWIPE patch data: $\boldsymbol{d}_\text{LSPE}$. The power spectra are estimated by pure-$B$ PCL method (in blue), and map combination method using filtered Planck-like simulations with FFP10 effective noise level ($\boldsymbol{\widehat d}_\text{Planck2}$), and scalar PCL (our method for case VI, in red). The black curve represent the input CMB $B$-mode power spectrum ($r$ = 0.05). The grey region shows the optimal error bounds of eq \ref{eq:cosmic_var}. The red region shows scalar PCL errors obtained from simulations without $E$-to-$B$ leakage. The dashed cyan curve shows the noise contribution to the optimal error limits.}
    \label{fig:Planck-LSPE}
\end{figure}
\begin{figure}[tbp]
\centering
    \subfloat{\includegraphics[width=0.5\textwidth]{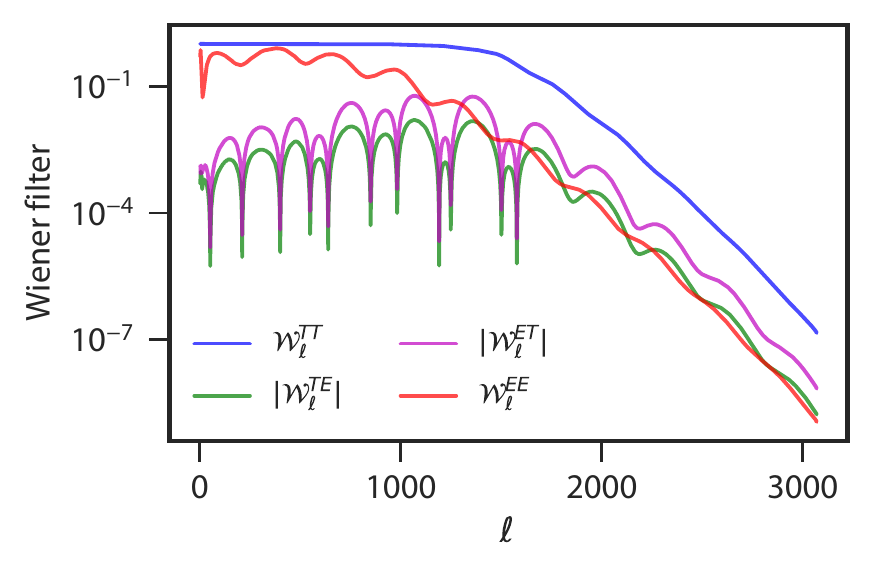}}
    \subfloat{\includegraphics[width=0.5\textwidth]{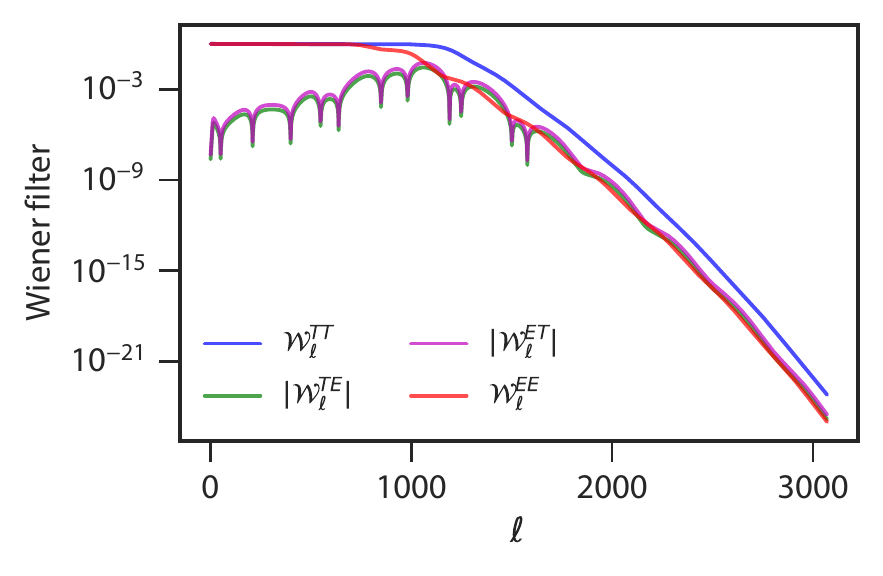}}\\
    \caption{\label{fig:wiener_filter} Wiener filter matrix components plotted against multipoles. Left: For Planck 100-143-217 GHz effective noise power spectra. Right: For LiteBIRD 68 to 235 GHz channels combined effective noise power spectra.}
\end{figure}

We use a multivariate Wiener filter to build the minimum-variance $E$-mode map, using the information from CMB temperature along with the $E$-modes to minimize the $E$-mode map error. Using $T$ is particularly important when the data outside the patch comes from Planck observations, as the CMB temperature map is signal dominated, and thus it helps improve $E$-mode signal reconstruction as compared to using polarization data alone. The multivariate Wiener filter can be written as \cite{Delabrouille2009}:
\begin{equation}
    \boldsymbol{\mathcal{W}} = \boldsymbol{C}\boldsymbol{I}^\dagger \left[\boldsymbol{I}\boldsymbol{C}\boldsymbol{I}^\dagger + \boldsymbol{N} \right]^{-1},
\end{equation}
where:
\begin{align}
    \boldsymbol{\mathcal{W}} = \begin{bmatrix}
        \mathcal{W}^{TT}_\ell & \mathcal{W}^{TE}_\ell\\
        \mathcal{W}^{ET}_\ell & \mathcal{W}^{EE}_\ell
    \end{bmatrix}; \quad & \quad \boldsymbol{C} = \begin{bmatrix}
        C_\ell^{TT}B_{T,\ell}^2 & C_\ell^{TE}B_{T,\ell}B_{E,\ell}\\
        C_\ell^{TE}B_{T,\ell}B_{E,\ell} & C_\ell^{EE}B_{E,\ell}^2
    \end{bmatrix}; \quad & \quad \boldsymbol{N} = \begin{bmatrix}
        N_\ell^{TT} & 0\\
        0 & N_\ell^{EE}.
    \end{bmatrix}
    \nonumber
\end{align}
Here $\boldsymbol{I}$ is the $2\times 2$ identity matrix. We plot the different elements of the multivariate Wiener filter in figure \ref{fig:wiener_filter} for both Planck and LiteBIRD effective noise power spectra. We can model our observed data as $d_{X,\ell m} = s_{X,\ell m} + n_{X,\ell m}$. Then the Wiener filtered satellite mission data is: $\boldsymbol{\widehat d}_{\ell m} = \boldsymbol{\mathcal{W}}\boldsymbol{d}_{\ell m}$, where $\boldsymbol{d}_{\ell m} = \begin{bmatrix}
    d_{T,\ell m} \\ d_{E, \ell m}
\end{bmatrix}$.
The Wiener filtered $E$-mode  spherical harmonic coefficients $\widehat d_{E, \ell m}$ can be written as:
\begin{equation}
    \widehat d_{E, \ell m} = \mathcal{W}^{ET}_\ell d_{T, \ell m} + \mathcal{W}^{EE}_\ell d_{E, \ell m}.
    \label{eq:WFed-E}
\end{equation}
We use the filtered $\widehat d_{E, \ell m}$, and with $d_{B,\ell m}$ set to zero, we re-synthesize the $QU$ map ($\boldsymbol{\widehat d}_\text{sat}$) with only the Wiener filtered $E$-mode signal and no $B$-mode signal. This map is then smoothed with a Gaussian beam of the same resolution as that of the ground-based experiment. This gives a minimum-variance $E$-mode-only $QU$ map over the full sky. 

\begin{figure}[tbp]
\centering 
\subfloat{\includegraphics[width=.25\textwidth]{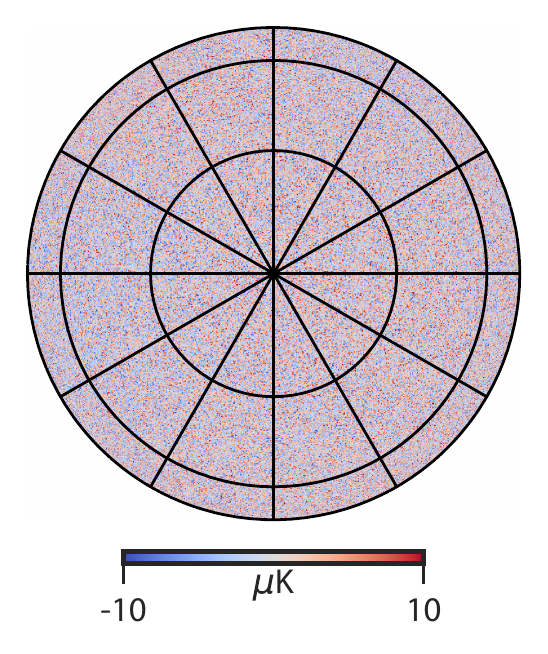}}
\subfloat{\includegraphics[width=.25\textwidth]{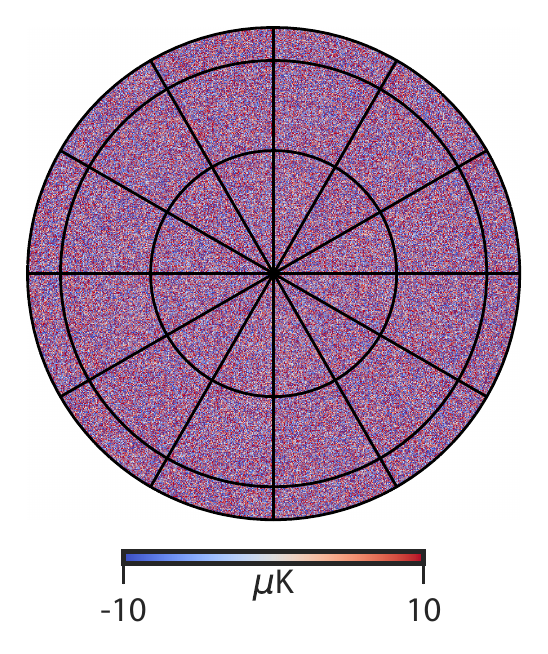}}
\subfloat{\includegraphics[width=.25\textwidth]{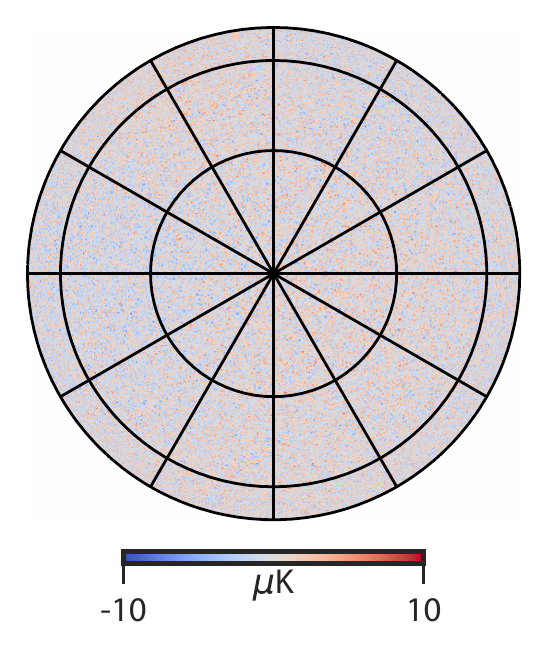}}
\subfloat{\includegraphics[width=.25\textwidth]{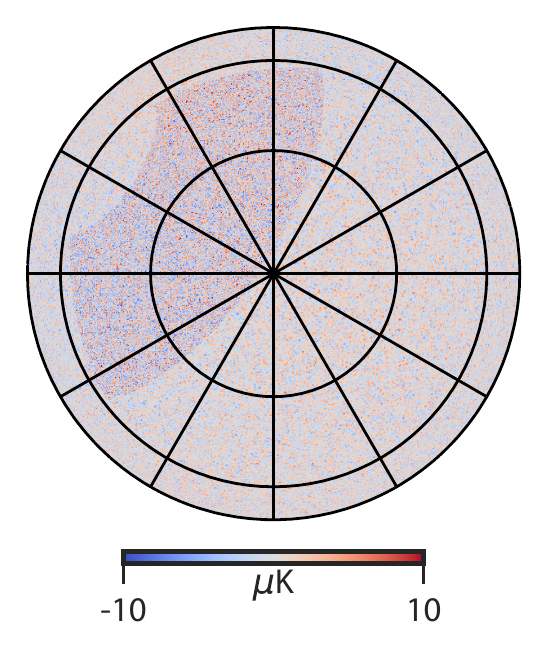}}
\caption{\label{fig:WFed-maps} Effect of Wiener filtering and map combination. Far left: Planck CMB $E$ map $\boldsymbol{s}_\text{Planck}$, Center left: Planck isotropized noise simulation from effective power spectrum $\boldsymbol{n}_\text{Planck2}$, Center right: Wiener filtered Planck observation $\boldsymbol{\widehat d}_\text{Planck2}$, and Far right: $E$ map of the combination $\boldsymbol{d}_\text{com}$ for $\boldsymbol{d}_\text{Planck2}$ and $\boldsymbol{d}_\text{patch1}.$ }
\end{figure}

\begin{figure}[tbp]
\centering 
\subfloat{\includegraphics[width=.33\textwidth]{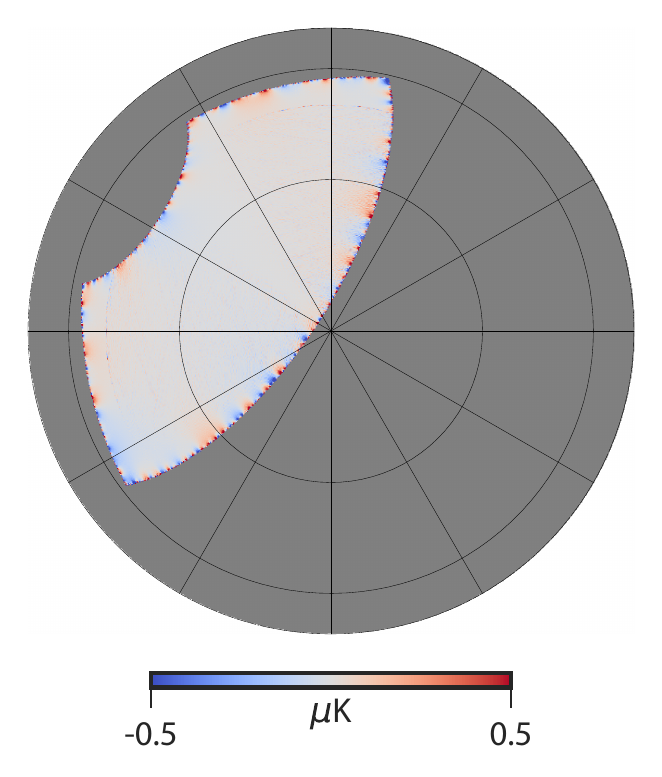}}
\subfloat{\includegraphics[width=.33\textwidth]{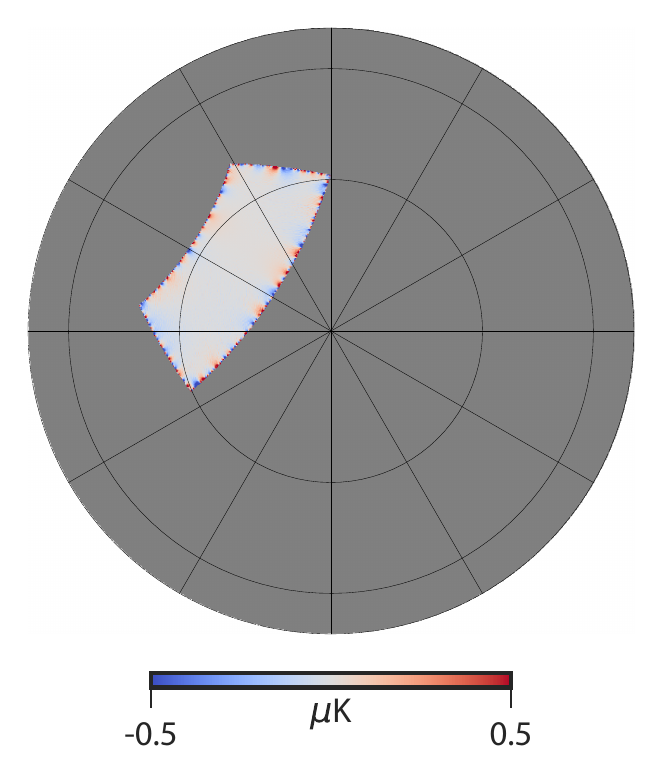}}
\subfloat{\includegraphics[width=.33\textwidth]{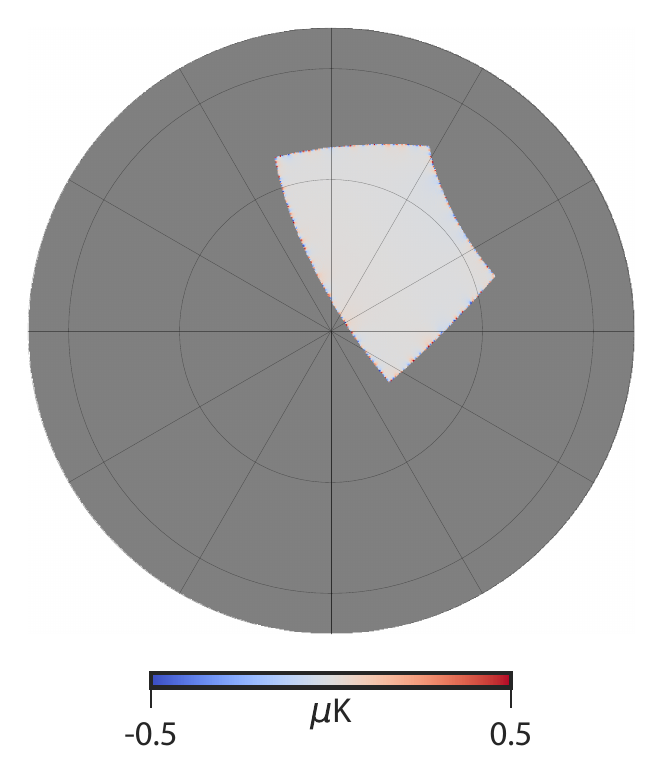}} \\
\subfloat{\includegraphics[width=.49\textwidth]{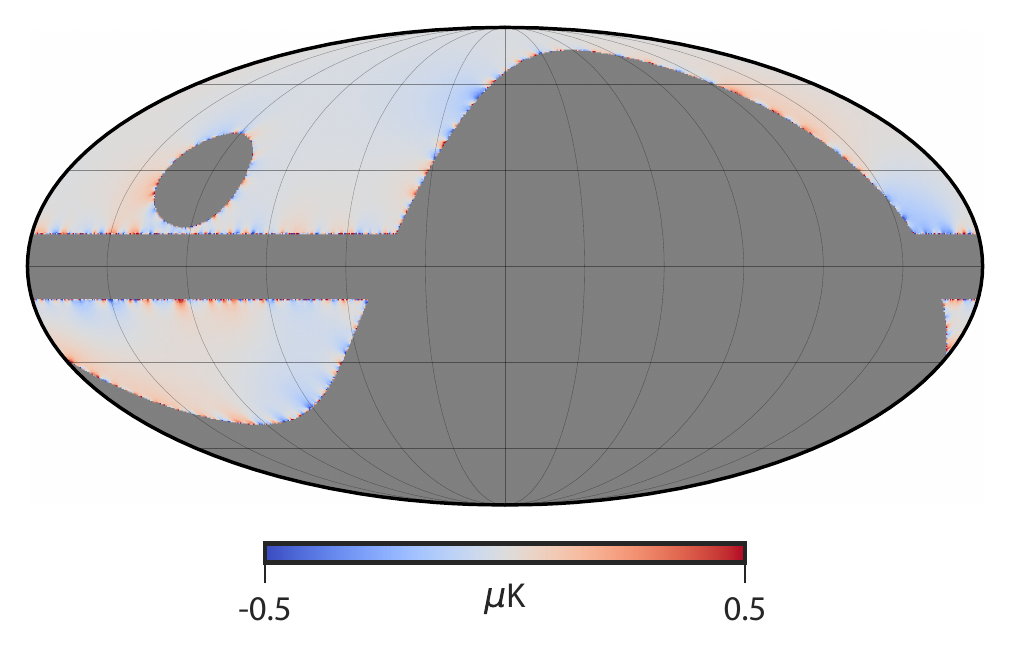}}
\caption{\label{fig:wCOM_leakage-maps} $E$-$B$ leakage with $E$-mode combination, for the three sky patches considered here. Top left: north sky patch 1 with  $\boldsymbol{d}_\text{patch1}$ - $\boldsymbol{\widehat d}_\text{Planck1}$ combination; Top center: north sky patch 2 with  $\boldsymbol{d}_\text{patch2}$ - $\boldsymbol{\widehat d}_\text{Planck1}$ combination; Top right: south sky CMB-S4 patch with  $\boldsymbol{d}_\text{CMB-S4}$ - $\boldsymbol{\widehat d}_\text{LiteBIRD}$ combination; Bottom: LSPE-SWIPE patch with $\boldsymbol{d}_\text{LSPE}$ - $\boldsymbol{\widehat d}_\text{Planck2}$ combination. The leakage is mostly concentrated at the edge of the patch.}
\end{figure}

The final step of the process is to combine the $QU$ map of the ground based (or balloon borne) experiment, $\boldsymbol{d}_\text{grd}$, with the full sky filtered $E$-mode-only $QU$ map we produced above, $\boldsymbol{\widehat d}_\text{sat}$. We combine the two sets of data by filling the outside of the ground observation patch ($\boldsymbol{W}_\text{grd}$) with the filtered satellite experiment data: 
\begin{equation}
    \boldsymbol{d}_\text{com} = \boldsymbol{\widehat d}_\text{sat}\left(1 - \boldsymbol{W}_\text{grd}\right) + \boldsymbol{d}_\text{grd}\boldsymbol{W}_\text{grd}.
\end{equation}
The unfiltered, simulated, $\boldsymbol{d}_\text{Planck2}$ $E$-mode  CMB signal, $E$-mode noise in the northern hemisphere are shown in the left two sub-figures of figure \ref{fig:WFed-maps}. The Wiener filtered $E$-mode observation for this case and the combined $E$-mode map from $\boldsymbol{d}_\text{com}$, when it is combined with $\boldsymbol{d}_\text{patch1}$, are shown in the two right sub-figures of figure \ref{fig:WFed-maps}. We can see that while the Wiener filter is highly effective in suppressing the $E$-mode noise in the simulated Planck maps, it also removes some part of the $E$-mode CMB signal. 
Therefore the final combined $E$-mode map shows a discontinuity between the observed patch and outside, seen in the right panel of figure \ref{fig:WFed-maps}. In figure \ref{fig:wCOM_leakage-maps} we show the leakage maps for our method for the three ground patches for the same CMB and noise realizations as those shown in figure \ref{fig:leakage-maps}. We can clearly see a significant reduction in the $E$-$B$ leakage with our method. The residual leakage is now mostly concentrated along the edges of the observation patch, occurring from the discontinuity in the $E$-mode signal at the edge of the patch. The level of residual leakage depends largely on the noise level of the full sky data. 


\begin{figure}[tbp]
\centering 
\subfloat{\includegraphics[width=.33\textwidth]{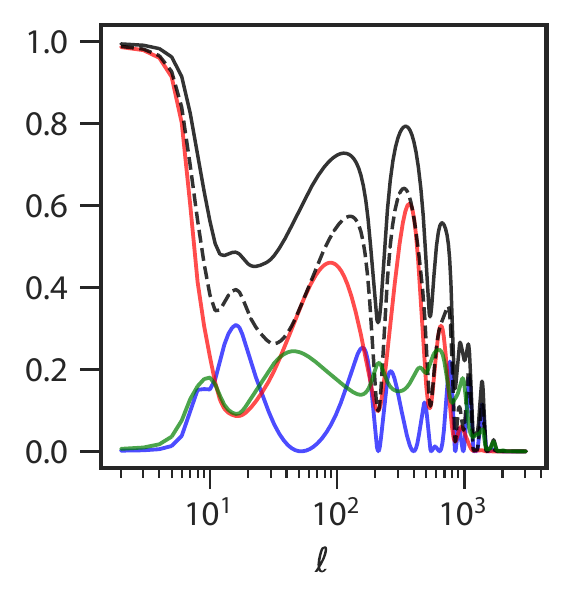}}
\subfloat{\includegraphics[width=.33\textwidth]{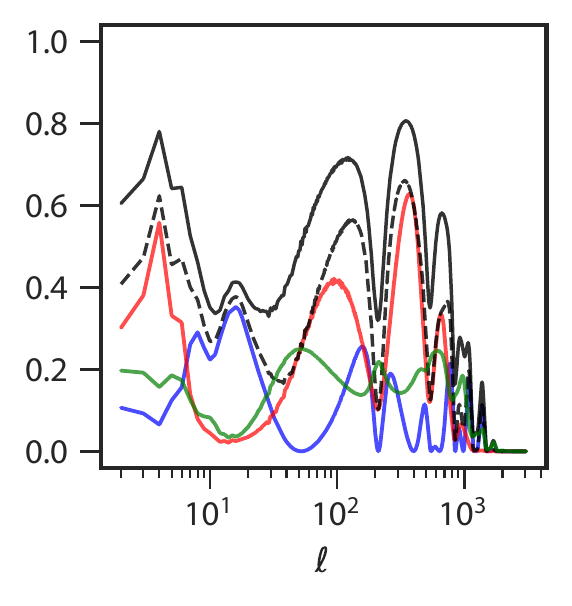}}
\subfloat{\includegraphics[width=.33\textwidth]{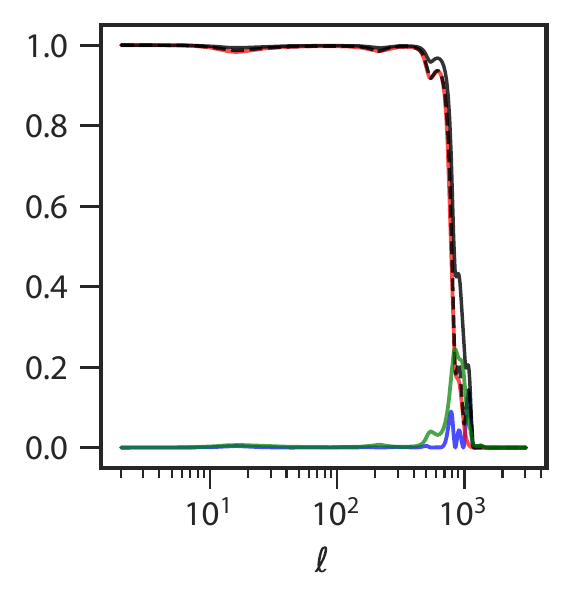}}
\caption{\label{fig:WF-contributions} Plots showing power ratio $r_p$ (solid black line), reconstruction ratio $r_\text{rec}$ (dashed black line), projected $E$-mode ratio $r_{\widehat E}$ (in red), projected $T$-mode ratio $r_{\widehat T}$ (in blue) and projected noise ratio $r_{\widehat n}$ (in green). Left: Planck with 50 $\mu$K-arcmin noise; centre: Planck with isotropized effective FFP10 noise; right: LiteBIRD case.}
\end{figure}

If our filtered $E$-mode map closely resembles the actual $E$-mode signal, we should have better leakage reduction. To have an idea of how much of the $E$-mode signal is correctly reconstructed in our $E$-mode map, we will construct a few important quantities. First, we compare the power spectrum of the filtered $E$ map to the $E$-mode signal by computing the power ratio, which we define as: 
\begin{equation}
   r_p = \frac{\langle \widehat d_{E, \ell m} \widehat d_{E, \ell' m'}^*\rangle}{C^{EE}_\ell B^2_{E,\ell}} 
\end{equation}
The contribution of the $E$-mode and $T$-mode projected signal in the filtered map may be studied with:
\begin{align}
    r_{\widehat E} = \frac{\langle \widehat d_{E, \ell m} \mathcal{W}^{EE}_{\ell'} s_{E, \ell' m'}^*\rangle}{C^{EE}_\ell B^2_{E,\ell}} \qquad r_{\widehat T} = \frac{\langle \widehat d_{E, \ell m} \mathcal{W}^{ET}_{\ell'} s_{T, \ell' m'}^*\rangle}{C^{EE}_\ell B^2_{E,\ell}}
\end{align}
The total $E$-mode signal reconstructed from the noisy $E$-mode and $T$-mode information can be studied with the reconstruction ratio:
\begin{equation}
    r_\text{rec} = \frac{1}{C^{EE}_\ell B^2_{E,\ell}}\left[\langle \widehat d_{E, \ell m} \mathcal{W}^{EE}_{\ell'} s_{E, \ell' m'}^*\rangle + \langle \widehat d_{E, \ell m} \mathcal{W}^{ET}_{\ell'} s_{T, \ell' m'}^*\rangle \right].
\end{equation}
Finally, while the Wiener filter reconstructs the $E$-mode signal from noisy $E$-mode and $T$-mode data, there is a fraction of the initial noise that it will fail to remove in the filtered $E$-mode  maps. To study the level of residual noise in the filtered maps we can construct the projected noise ratio:
\begin{equation}
    r_{\widehat n} = \frac{\langle \widehat d_{E, \ell m} \widehat n_{E, \ell' m'}^*\rangle}{C^{EE}_\ell B^2_{E,\ell}}
\end{equation}
In figure \ref{fig:WF-contributions} we show the plot of the different ratios discussed here. The relations for the correlations used in calculating these ratios is given in appendix \ref{sec:appendix}. We can see from figure \ref{fig:WF-contributions} that the reconstruction ratio for either of the two Planck sky noise cases is not ideal. Hence, only a fraction of the $E$-mode discontinuity at the patch edges can be corrected with our method. For most of the multipole range $r_\text{rec}$ is below 0.6. The signal reconstruction for $\boldsymbol{d}_\text{Planck2}$ case is the worst of the three full sky maps considered in our work, especially on large scales (it becomes equivalent to the $\boldsymbol{d}_\text{Planck1}$ case above $\ell=70$). This behaviour is explained by the noise level in the various cases considered here. The low $E$-mode signal reconstruction (low $r_\text{rec}$) problem for all the filtered Planck-like simulations arises because the noise level in Planck $E$-mode maps is high, and the Wiener filter aggressively suppresses the noise at the cost of removing the signal from the filtered maps. The importance of the multivariate Wiener filter can also be seen from figure \ref{fig:WF-contributions}. For $\boldsymbol{d}_\text{Planck1}$ and especially $\boldsymbol{d}_\text{Planck2}$ we find that around $\ell=20$ the signal dominated $T$ modes contribute more than the noisy $E$ mode towards the filtered map. It thus helps to utilize the higher quality $T$-mode maps in the signal reconstruction. The $\boldsymbol{d}_\text{LiteBIRD}$ reconstruction is totally signal dominated as for most of the multipole range $r_\text{rec}$ is above 0.9. Finally, we note that the LiteBIRD $E$ modes are signal dominated up to $\ell \simeq 700$, therefore the $T$-mode information is not used in the signal reconstruction except on the smallest scales, when the $E$-mode noise becomes comparable to the level of the $E$-mode signal. Another point to note from figure \ref{fig:WF-contributions} is the projected noise ratio for either of the Planck cases is around 0.15 for most of the multipole range. This indicates the residual noise level in the filtered Planck maps. However, for signal dominated LiteBIRD the noise ratio is negligible till high multipoles.

\begin{figure}[tbp]
\centering 
\subfloat[$\boldsymbol{d}_\text{patch1}$ - $\boldsymbol{\widehat d}_\text{Planck1}$]{\includegraphics[width=.5\textwidth]{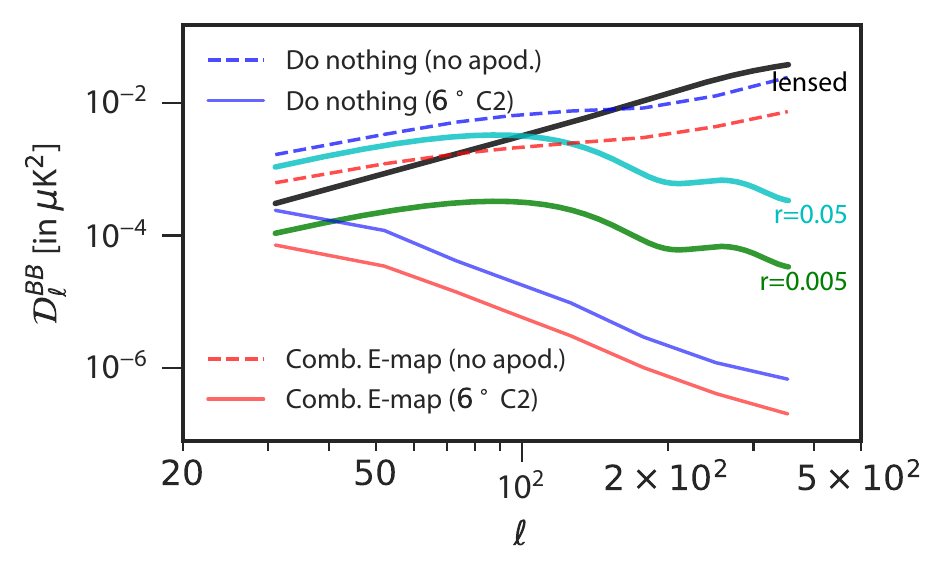}}
\subfloat[$\boldsymbol{d}_\text{patch2}$ - $\boldsymbol{\widehat d}_\text{Planck1}$]{\includegraphics[width=.5\textwidth]{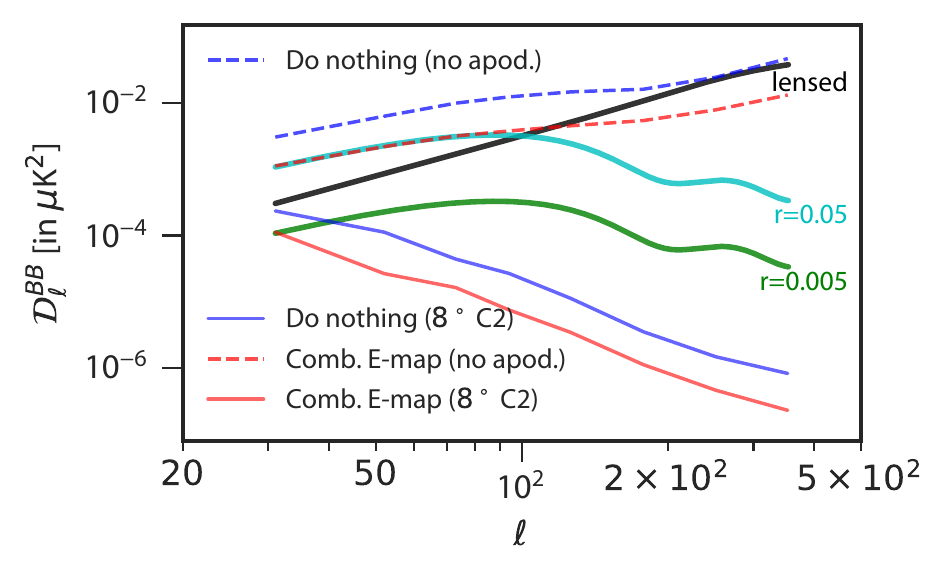}}\\
\subfloat[$\boldsymbol{d}_\text{patch1}$ - $\boldsymbol{\widehat d}_\text{ Planck2}$]{\includegraphics[width=.5\textwidth]{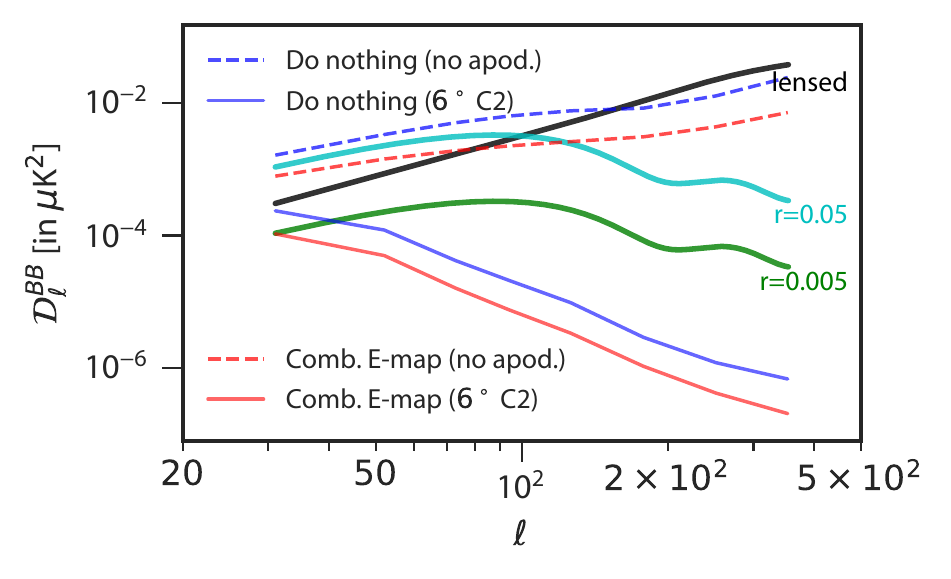}}
\subfloat[$\boldsymbol{d}_\text{patch2}$ - $\boldsymbol{\widehat d}_\text{ Planck2}$]{\includegraphics[width=.5\textwidth]{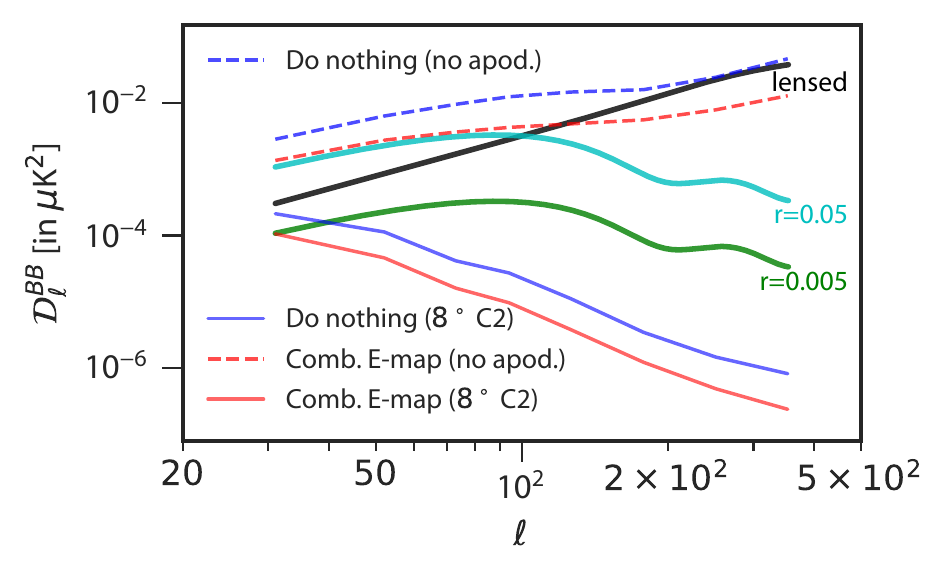}}\\
\subfloat[$\boldsymbol{d}_\text{CMB-S4}$ - $\boldsymbol{\widehat d}_\text{ LiteBIRD}$]{\includegraphics[width=.5\textwidth]{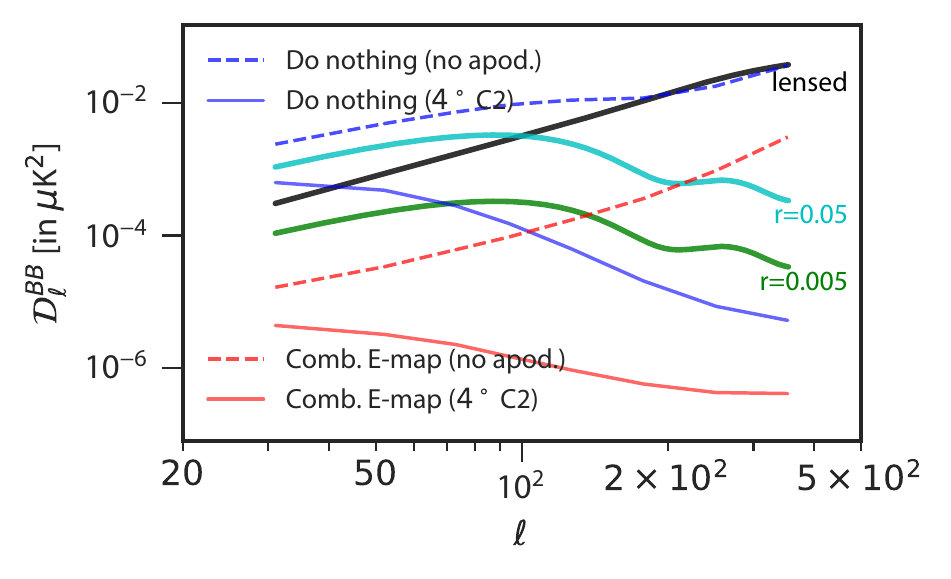}}
\subfloat[$\boldsymbol{d}_\text{LSPE}$ - $\boldsymbol{\widehat d}_\text{ Planck2}$]{\includegraphics[width=.48\textwidth]{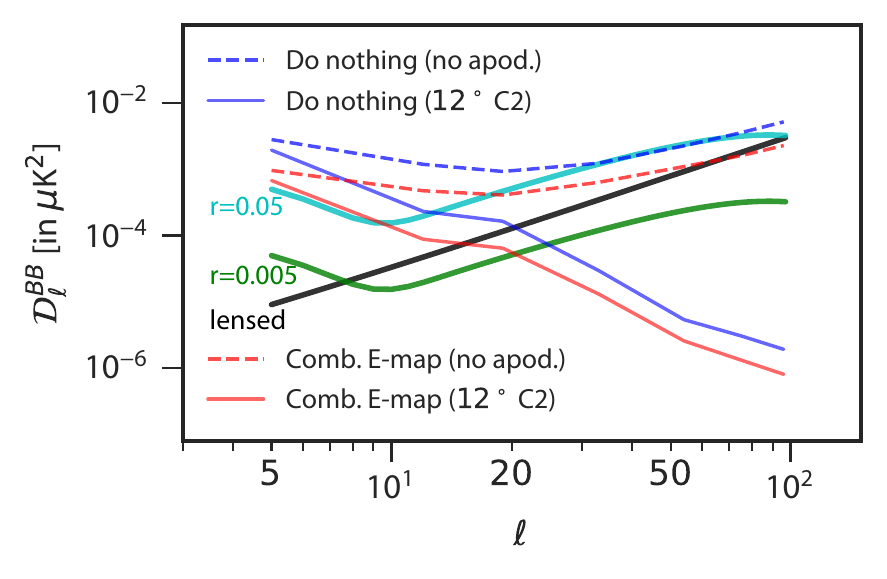}}\\
\caption{\label{fig:leakage_spec} Plots showing the leakage spectra for the five different cases of combined $E$-map method and doing nothing. The cosmological signals in the plots are: lensing $B$-modes (in black), $r=0.05$ primordial $B$-modes (in cyan) and $r = 0.005$ primordial $B$-modes (in green). The leakage spectrum for doing no $E$-$B$ leakage control is shown in blue, while the leakage spectrum for our combined $E$-map method is shown in red. The dashed lines are the leakage spectra obtained without apodization. The leakage spectra obtained with apodization is shown with solid lines. The apodization length is mentioned in the legends of individual figures.}
\end{figure}

We study the level of residual leakage by obtaining the power spectra of the leakage maps. Since we did not have any input $B$-mode signal or noise, we call this the leakage spectrum because it arises only due the $E$-to-$B$ leakage due the incompleteness of the sky.  We compare the leakage spectrum for our $E$-map combination method (for maps like those shown in figure \ref{fig:wCOM_leakage-maps}) to the case where we do nothing to control the $E$-to-$B$ leakage (like maps shown in figure \ref{fig:leakage-maps}). In figure \ref{fig:leakage_spec} we show the mean leakage spectra for 300 simulations (800 simulations for LSPE patch). The plots show results both with, and without apodization. We use the same apodization that we use for analysis of the $B$-mode  power spectrum results for each case. From figure \ref{fig:leakage_spec} we see that the leakage is reduced when we combine the $E$-mode maps and that the residual leakage is along the edges. In all cases, when using apodization, the leakage spectra for the combined $E$-map method is well below the total $B$-mode power spectrum. We can also relate the performance to the performance of our Wiener filtering by consulting figure \ref{fig:WF-contributions}. The better the signal reconstruction in the filtered $E$-mode maps, the more suppressed will be the residual leakage. In the filtered full sky maps outside the observation patch, a fraction ($1-r_\text{rec}$) of the $E$-mode signal is missing. It is this mismatch in the $E$-mode  signal levels that results in the residual leakage. Therefore, when $r_\text{rec}$ is high, as will be the case with LiteBIRD data, the residual leakage is negligible. The performance with Planck maps is not quite ideal (because the Planck $E$-mode map is not strongly signal-dominated), but the $E$-to-$B$ leakage reduction achieved by our method is still significant (a factor of about 3 in the leakage level). For both patches 1 and 2, when combining with Planck signal, the residual leakage with masking is below the total $B$-mode power spectrum. For $\ell < 20$, on the LSPE patch, while the combined $E$-map method reduces the leakage significantly, there is more optimization required to further suppress the leakage on the largest scales. 
Part of the relatively large residual leakage can be attributed to the low signal-to-noise ratio for Planck on these large angular scales. However, the level of $E$-to-$B$ leakage also depends on the number of useful independent modes of observation impacted by the boundary, as compared to the number of independent modes effectively measured inside the patch, away from the boundary. In this respect, the LSPE sky coverage with respect to the measurement of the largest scales is the most complicated patch that we assume in this work. Most of the low-$\ell$ independent structures are cut by the edge of the observed region, and the shape complexity and the large boundary of the patch contributes significantly to the leakage problem.

We obtain the $B$-mode power spectrum from our method, for the following six different partial sky and filtered full sky observation pairs: I. $\boldsymbol{d}_\text{patch1}$ - $\boldsymbol{\widehat d}_\text{Planck1}$, II. $\boldsymbol{d}_\text{patch2}$ - $\boldsymbol{\widehat d}_\text{Planck1}$, III. $\boldsymbol{d}_\text{patch1}$ - $\boldsymbol{\widehat d}_\text{Planck2}$, IV. $\boldsymbol{d}_\text{patch2}$ - $\boldsymbol{\widehat d}_\text{Planck2}$, V. $\boldsymbol{d}_\text{CMB-S4}$ - $\boldsymbol{\widehat d}_\text{LiteBIRD}$, VI. $\boldsymbol{d}_\text{LSPE}$ - $\boldsymbol{\widehat d}_\text{Planck2}$. For each of them we first get the $B$-mode map from the combined $QU$ map, $\boldsymbol{d}_\text{com}$. This is a $B$-mode map with $E$-to-$B$ leakage reduced. Note that the combination of the filtered full sky satellite data in the outside patch does not add any $B$-mode  noise or signal to the combined $B$ map, as we had set the $B$ modes in our filtered $QU$ map to zero. Any additional contribution to this map would come from residual leakage discussed above. The $B$-mode signal is present only in the observation patch, hence we will mask it with an C2 apodized mask \cite{Grain2009}, and proceed with power spectrum estimation by the scalar pseudo-$C_\ell$ method of equation \eqref{eq:scalar_mix}. The apodization helps in reducing the influence of the residual leakage at the edges of the patch. We have implemented the scalar pseudo-$C_\ell$ with NaMaster python package. We plot the mean and standard deviation for 300 random simulations for the cases I through V and are shown in figures \ref{fig:fsky8pc_50-10} to \ref{fig:liteBIRD-CMB-S4}. The result for case VI are computed from 800 simulations and is shown in figure \ref{fig:Planck-LSPE}. For cases I to V, we can clearly see from all the cases plotted here that our new method gives us unbiased estimate of the $B$-mode  power spectra with near-optimal error bars. The performance slightly deteriorates with smaller $f_\text{sky}$, but remains close to the theoretical minimum for all the cases studied in this paper. On the largest angular scales for $\ell < 20$, we find from case VI that our method gives nearly unbiased estimates for the power spectrum. 
It is known \cite{Efstathiou2004, Challinor2005} that the PCL errors at low-$\ell$s are more complicated than the optimal errors of equation \ref{eq:cosmic_var}. To test this we simulated LSPE-like maps with $E$-mode signal and noise set to zero. This ensured no $E$-to-$B$ leakage in the test. We then obtained $B$-mode map for the LSPE patch and used scalar PCL estimator to obtain the power spectra for these maps. The variance of these $B$-mode power spectra estimates gives the error from the PCL estimator alone. From figure \ref{fig:Planck-LSPE} we can clearly see that the PCL error even in absence of any $E$-to-$B$ leakage is larger than the optimal errors of \ref{eq:cosmic_var}. The error bars for the combined $E$-map method are largely consistent with the scalar PCL error band obtained from simulations. This shows that the large errors at low multipoles are mostly due to the PCL estimator, while some of this may also have resulted from the low signal-to-noise of the Planck $E$ modes. 

\section{Discussion} \label{sec:discussion}
We have seen in the previous section that the $E$-map combination method introduced here performs adequately well for the cases presented in this work. Even with noisy full sky $E$-mode data we are able to suppress the $E$-to-$B$ leakage sufficiently. A comparison with the standard method and the pure-$B$ method for $\boldsymbol{d}_\text{patch1}$ - $\boldsymbol{\widehat d}_\text{Planck1}$ and $\boldsymbol{d}_\text{patch2}$ - $\boldsymbol{\widehat d}_\text{Planck1}$ (shown in figures \ref{fig:fsky8pc_50-10} and \ref{fig:fsky2pc_50-5} respectively), shows that our method outperforms both the standard and the pure-$B$ methods. While pure-$B$ method is a close second, it doesn't have optimal error bars for the first two multipole bins in the cases considered here. The standard method with $E$-$B$ mixing inversion is the worst performer of the three methods. We also note that our $E$-map combination method has computation times that are comparable to the pure-$B$ method. Therefore, the combined $E$-map PCL method is a fast method for power spectrum estimation.

The combined $E$-map method performs significantly better than the pure-$B$ method for the smaller sky patch and for $r=0$ case. For the $\boldsymbol{d}_\text{patch1}$ - $\boldsymbol{\widehat d}_\text{Planck2}$ and $\boldsymbol{d}_\text{patch2}$ - $\boldsymbol{\widehat d}_\text{Planck2}$ cases (shown in figure \ref{fig:fsky8pc_FFP10-10} and \ref{fig:fsky2pc_FFP10-5}) we find that, even with isotropized Planck effective noise levels, our method outperforms the pure-$B$ method with near optimal error bars. 

In the $\boldsymbol{d}_\text{CMB-S4}$ - $\boldsymbol{\widehat d}_\text{LiteBIRD}$ case (shown in figure \ref{fig:liteBIRD-CMB-S4}) the pure-$B$ and our method perform comparably well, but our method has a slightly smaller (optimal) error bar in the first bin. 

In the $\boldsymbol{d}_\text{LSPE}$ - $\boldsymbol{\widehat d}_\text{Planck2}$ case we see substantial improvement with our combined $E$-map method over the pure-$B$ PCL method. We are being impaired by the use of the PCL estimator for the low multipoles. We anticipate that using a quadratic maximum likelihood estimator instead of the PCL estimator can alleviate much of the problem \cite{Efstathiou2004}. This case also highlights that some more work is needed to further optimize the performance of our method at low multipoles with the Planck data. 
We will postpone these changes and optimizations to future work.

When full-sky $E$-mode maps with high signal-to-noise ratio will be available, $E$-to-$B$ leakage due to incomplete sky analyses with future ground-based experiments can trivially be solved using our proposed approach. However, even within the current constraints, all the results presented in this paper highlight the usefulness of our method and it's excellent performance as compared to two of the most common and standard methods used in the community. 

It is clear from the discussion so far that the noise level of the full sky $E$-mode observations is critical for the good performance of the new method. The noise level of currently available full sky data is constrained by the already completed Planck mission observations. In the near future however, ground-based observations can also improve on the measurement of $E$ modes on a sky region significantly larger than the deep sky patch used to search for primordial $B$ modes. These observations, if they surround the deep patch, can be combined with Planck for a better $E$-mode maps and yet reduced $E$ to $B$ leakage for primordial $B$-mode analysis. This has implications on the choice of a scan strategy design for sub-orbital CMB polarization experiments in the near future, as additional shallower observations around a deep $B$-mode observation patch, with a signal-to-noise ratio sufficient to observe the $E$-modes only, can help make a smooth merging of the ground-based $E$-mode signal and existing satellite $E$-mode signals. 

The results we have obtained in this work show that reducing $E$-to-$B$ leakage by completing the $E$-mode signal is potentially the ultimate method for $E$-to-$B$ leakage reduction due to partial sky observations. In practice, this approach will have to take into account additional complications such as inhomogeneous and correlated noise, timestream-level filtering, foreground residuals, etc. Although there is no conceptual showstopper for doing so, we postpone these refinements to future study. 

\section{Conclusion}
In this work we have proposed a new method of reducing the $E$-to-$B$ leakage, by completing the $E$-mode information in the the area outside the observation patch of a ground-based survey. Our results show that this method gives unbiased estimates for the $B$-mode power spectra with near-optimal error for $\ell>20$. We find that the performance depends on the noise level of the full sky data. However, in all the cases considered the new method outperforms the pure-$B$ method. Though the $E$-map combination method needs further testing and adjustments for realistic cases, the initial results shown here are promising, and suggest that the complications of partial sky $E$-$B$ ambiguity for the detection of primordial $B$ modes are about to be optimally solved.

\appendix
\section{Some important relations} \label{sec:appendix}
We can use equation \eqref{eq:WFed-E} to calculate correlations used for constructing the different ratios defined in section \ref{ssec:method}. The power spectrum of the Wiener filtered $E$-mode map is given by:
\begin{align}
   \langle \widehat d_{E, \ell m} \widehat d_{E, \ell' m'}^*\rangle = (\mathcal{W}^{ET}_\ell)^2 \left[C^{TT}_\ell B^2_{T,\ell} + N^{TT}_\ell \right] & + (\mathcal{W}^{EE}_\ell)^2 \left[C^{EE}_\ell B^2_{P,\ell} + N^{EE}_\ell \right] \nonumber \\ & + 2 \mathcal{W}^{ET}_\ell \mathcal{W}^{EE}_\ell C^{TE}_\ell B^2_{TP, \ell}
\end{align}
The contribution of the projected $T$-mode signal to the Wiener filtered $E$-mode power is given as:
\begin{equation}
   \langle \widehat d_{E, \ell m} \mathcal{W}^{ET}_{\ell'} s_{T, \ell' m'}^*\rangle = (\mathcal{W}^{ET}_\ell)^2 C^{TT}_\ell B^2_{T,\ell} + \mathcal{W}^{ET}_\ell \mathcal{W}^{EE}_\ell C^{TE}_\ell B^2_{TP, \ell} 
\end{equation}
The contribution of the projected $E$-mode signal is given as:
\begin{equation}
   \langle \widehat d_{E, \ell m} \mathcal{W}^{EE}_{\ell'} s_{E, \ell' m'}^*\rangle = (\mathcal{W}^{EE}_\ell)^2 C^{EE}_\ell B^2_{P, \ell} + \mathcal{W}^{ET}_\ell \mathcal{W}^{EE}_\ell C^{TE}_\ell B^2_{TP, \ell}
\end{equation}
The total projected noise contribution to the Wiener filtered $E$-mode maps is:
\begin{equation}
   \langle \widehat d_{E, \ell m} \widehat n_{E, \ell' m'}^*\rangle = (\mathcal{W}^{ET}_\ell)^2 N^{TT}_\ell + (\mathcal{W}^{EE}_\ell)^2 N^{EE}_\ell
\end{equation}

\acknowledgments
Some of the results in this paper have been derived using the HEALPix \cite{Gorski2005} package. This work is supported by NSFC Grants No. 11903030, No. 11773028, No. 11633001, No. 11653002, No. 11421303. 

\bibliographystyle{JHEP}
\bibliography{biblio}

\end{document}